\documentclass[showpacs,prb,twocolumn]{revtex4}
\usepackage{amsfonts}
\usepackage{amsmath}
\usepackage{amssymb}
\usepackage[final,dvips]{graphicx}%
\begin{document}
\title{Vortex-chain phases in layered superconductors}
\author{A. E. Koshelev}
\affiliation{Materials Science Division, Argonne National
Laboratory, Argonne, Illinois 60439}

\date{\today}

\begin{abstract}
Layered superconductors in tilted magnetic field have a very rich
spectrum of vortex lattice configurations. In the presence of
in-plane magnetic field, a small c-axis field penetrates in the
form of isolated vortex chains. The structure of a single chain is
mainly determined by the ratio of the London [$\lambda$] and
Josephson [$\lambda_{J}$] lengths, $\alpha= \lambda/\lambda_{J}$.
At large $\alpha$ the chain is composed of tilted vortices [tilted
chains] and at small $\alpha$ it consists of a crossing array of
Josephson vortices and pancake stacks [crossing chains]. We
studied the chain structures at intermediate $\alpha$'s and found
two types of behavior. (\textbf{I}) In the range $0.4 \lesssim
\alpha\lesssim 0.5$ a $c$-axis field first penetrates in the form
of pancake-stack chains located on Josephson vortices. Due to
attractive coupling between deformed stacks, their density jumps
from zero to a finite value. With further increase of the $c$-axis
field the chain structure smoothly evolves into modulated tilted
vortices and then transforms via a second-order phase transition,
into the tilted straight vortices. (\textbf{II}) In the range $0.5
\lesssim\alpha \lesssim 0.65$ a $c$-axis field first penetrates in
the form of kinks creating kinked tilted vortices. With increasing
the $c$-axis field this structure is replaced via a first-order
phase transition by the strongly deformed crossing chain. This
transition is accompanied by a large jump of pancake density.
Further evolution of the chain structure is similar to the higher
anisotropy scenario: it smoothly transforms back into the tilted
straight vortices.

\end{abstract}
\maketitle

\section{Introduction}

Layered superconductors have an amazingly rich phase diagram in
tilted magnetic field. In the presence of the in-plane field,
pancake vortices generated by the c-axis field
\cite{pancakes,ClemPRB91} can form a very large number of
different lattice configurations. Possible structures include the
kinked lattice, \cite{IvlevMPL1991,BLK,Feinberg93,Kinkwalls},
tilted vortex chains \cite{TiltedChains},  and crossing lattices
composed of sublattices of Josephson vortices (JVs) and
pancake-vortex stacks.\cite{BLK,CrossLatPRL99} In addition to
homogeneous lattices, phase-separated states may also exist such
as dense pancake-stack chains sitting on JVs and dilute lattice in
between \cite{Bolle91,Grig95,Huse,CrossLatPRL99} or coexisting
lattices with different orientations.\cite{coex-lat}
Even though considerable progress in this field has been made in
the last decade, the satisfactory understanding of the phase
diagram has not been achieved yet. All these phases probably do
realize in different materials and experimental conditions.
However finding ground states in tilted field occurs to be a
challenging theoretical task and it is even more difficult to
prove experimentally that a particular lattice configuration does
realize somewhere in the phase diagram.

The main source of richness of lattice structures in tilted field
is the existence of two very different kinds of interactions
between pancake vortices in different layers: magnetic and
Josephson interactions. The key parameter, which determines the
relative strength of these two interaction and plays a major role
in selecting the lattice structures, is the ratio of the two
fundamental lengths, the in-plane London penetration depth,
$\lambda \equiv\lambda_{ab}$, and Josephson length
$\lambda_{J}=\gamma s$, with $\gamma$ being the anisotropy
parameter and $s$ being the interlayer spacing, $\alpha=
\lambda/\lambda_{J}$. One can distinguish two limiting cases which
we refer to as ``extremely anisotropic'' case, $\alpha< 0.4$, and
``moderately anisotropic'' case $\alpha> 0.7$. Note that in our
terminology even ``moderately anisotropic'' superconductors may
have very large anisotropy factor, $\gamma\gg1$. Among known
atomically layered superconductors, only
Bi$_{2}$Sr$_{2}$CaCu$_{2}$O$_{x}$ (BSCCO) and related compounds
may belong to the ``extremely anisotropic'' family. Even in this
compound the parameter $\alpha$ is not smaller than $\sim0.25$ and
increases with temperature so that BSCCO typically becomes
``moderately anisotropic'' in the vicinity of transition
temperature.
\begin{figure}[ptb]
\begin{center}
\includegraphics[width=3.4in]{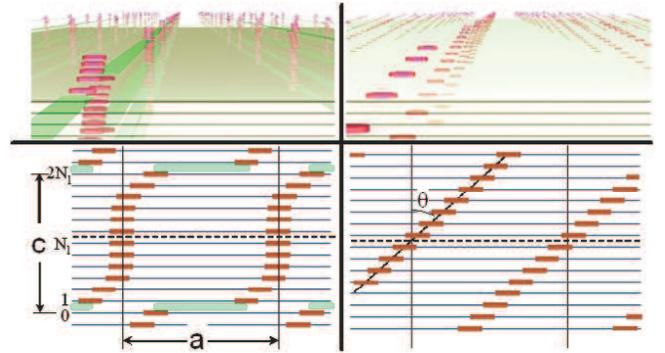}
\end{center}
\caption{Crossing (left) and tilted (right) vortex chains. Upper
pictures show three-dimensional views and lower pictures show the
structures of isolated chains.}%
\label{Fig-CrossTiltChain}%
\end{figure}

In a wide range of the in-plane fields (10-200 G) and at very
small c-axis fields (up to 1-2 G) the pancake stacks in layered
superconductors within a wide range of anisotropies are arranged
in chains, see Fig.\ \ref{Fig-CrossTiltChain}. An isolated chain
is a two-dimensional array of pancake vortices oriented
perpendicular to the layers. At somewhat higher $c$-axis fields
the chains are surrounded by the stripes of regular vortex
lattice.\cite{Bolle91,Grig95} The internal structure of an
isolated chain depends on the ratio $\alpha$ and it is relatively
simple in two limiting cases. At large $\alpha$ the chain is
composed of tilted pancake stacks (tilted chain, right column in
Fig.\ \ref{Fig-CrossTiltChain}) and at small $\alpha$ it consists
of crossing array of Josephson vortices and pancake stacks
(crossing chains, left column in Fig.\ \ref{Fig-CrossTiltChain}).
A very nontrivial and intriguing problem is how one structure
transforms into another in the region of intermediate $\alpha$. We
address this problem in this paper. We analytically and
numerically computed ground-state configurations in the isolated
vortex chain and found a surprisingly rich behavior. We found two
types of phase transitions. The first phase transition typically
takes place for the intermediate separations between pancake
stacks $a$, $a=[1-2]\lambda _{J}$, and rather wide range of the
ratio $\alpha$, $0.4 \lesssim \alpha\lesssim0.65$. For these
$\alpha$'s the ground state is given by the crossing chain in a
wide range of the pancake separations $a$. However, due to
attractive coupling between deformed pancake stacks
\cite{BuzdinPRL02}, the equilibrium separation can not exceed some
maximum value, which depends on the in-plane field and $\alpha$
and it is typically of the order of several $\lambda_{J}$. With
decreasing the pancake separation $a$, the crossing chain becomes
strongly deformed and smoothly transforms into the modulated
tilted vortices which then transform via a second-order phase
transition into the tilted straight vortices. We calculated
analytically the energies of two limiting chain configurations and
checked that numerics reproduces them. Comparing these energies,
we locate the transitional region in the phase space, where
strongly-deformed chains are realized.

Another phase transition is realized at very small densities of
pancake vortices and only when $\alpha$ exceeds a certain critical
value $\approx 0.5$ (exact criterion depends on the in-plane
magnetic field). In this case a small c-axis field penetrates in
the form of kinks. The kinked vortex lines forming tilted chains
are composed of pieces of Josephson vortices separated by
kinks.\cite{IvlevMPL1991,BLK,Feinberg93} If the kink energy is
only slightly smaller then the energy per pancake in a straight
pancake stack then at very small concentration of kinks, typically
at $a\approx[20-30]\lambda_{J}$, the kinked chains are replaced
with strongly deformed crossing chains via a first-order phase
transition. Due to the opposite signs of interactions (kinks repel
and deformed pancake stacks attract each other) this transition is
accompanied by a very large jump in the pancake density. With
further decrease of pancake separation the chain smoothly
transforms back to the tilted chain as it was described in the
previous paragraph.

Based on numerical exploration of the chain configurations, we construct the
chain phase diagrams for different ratios $\alpha$. As follows from the above
description, there are two types of phase diagrams in the region of
intermediate $\alpha$'s.
\begin{itemize}
\item In the range $0.4 \lesssim\alpha\lesssim 0.5$ a small
$c$-axis field first penetrates in the form of pancake-stack
chains located on Josephson vortices. Due to attractive coupling
between deformed stacks, their density jumps from zero to a finite
value. With further increase of the $c$ axis field the chain
structure first evolves into the modulated tilted vortices, which
then transforms via a second-order phase transition, into the
tilted straight vortices.%
\item In the range $0.5 \lesssim\alpha\lesssim0.65$ a small $c$
axis field first penetrates in the form of kinks creating kinked
tilted vortices. With increasing the $c$-axis field this structure
is replaced via a first-order phase transition by the chain of
pancake stacks, which are typically strongly deformed. This
transition is accompanied by a large jump of pancake density.
Further evolution of the chain structure is identical to the
smaller $\alpha$ scenario: the structure first transforms into
modulated tilted vortices and then, via a second-order phase
transition, into tilted straight vortices.
\end{itemize}
Note that the exact transition between the two types of behavior
depends also on the in-plane field. As interaction with the
Josephson vortices reduces the energy of the pancake stacks, the
larger in-plane field favors the first scenario.

Using numerical code developed for studying the chain structures,
we also investigated stability of an isolated crossing
configuration of the Josephson vortex and pancake stack. We found
that the crossing configuration becomes unstable at $\alpha\approx
0.69$. Above this value the magnetic coupling is not capable to
maintain stable configuration and the Josephson vortex tears the
stack apart. Nevertheless, the obtained stability range occurs to
be significantly broader than it was estimated from simple
considerations in Ref. \onlinecite{DodgsonPhysC02}. The reason is
that the strongly-deformed crossing configuration significantly
modifies the Josephson vortex which reduces forces pulling
pancakes apart. We found that the crossing energy increases
smoothly up to instability point. Perturbative calculation
\cite{CrossLatPRL99} gives accurate results for the crossing
configuration and its energy up to $\alpha \approx 0.35$.

Recently, the vortex chains in BSCCO at small concentrations of
pancakes have been studied by the scanning Hall probe microscopy
by Grigorenko \emph{et al.}\cite{GrigNat01} They observed that at
very small concentration of the pancakes the chains are
magnetically homogeneous and separate pancake stacks are not
resolved. When the external field exceeds some critical value of
the order of several Oersted, crystallites of the pancake stacks
are suddenly formed along the chain and the flux density in the
crystallites approximately ten times higher then the flux density
in the homogeneous chains. Our calculations provide consistent
interpretation for these observations. The magnetically
homogeneous chains are interpreted as kinked/tilted chains and
formation of crystallites can be attributed to the low-density
[kinked lines]-[crossing chains] first-order phase transition
(such interpretation has been proposed by Dodgson
\cite{DodgsonPRB02}). The observed large density jump also comes
out from the theory.

The evolution of the mixed chain+lattice state with increasing
temperature has been studied recently by Lorentz
microscopy.\cite{MatsudaSci02} It was observed that the pancake
stacks located in chains smear along the chain direction above
some field-dependent temperature while the pancake stacks outside
chains still remain well-defined. The continuous low-density phase
transition from crossing to tilted chain found and discussed in
this paper provides a very natural interpretation for this
observation.

The paper is organized as follows. In Section \ref{Sec-General} we
review general expressions for the chain energy. In Section
\ref{Sec-Analyt} we perform analytical calculations of the chain
energy for the two limiting cases: crossing and tilted chain. In
the next section \ref{Sec-TransRegion}, comparing energies for the
two limiting configurations, we estimate location of the
transition region. In section \ref{Sec-Attrac} we review
attractive interaction between deformed pancake stacks located on
Josephson vortices\cite{BuzdinPRL02} and derive general formulas
for determination of the maximum equilibrium separation between
the pancake stacks and the boundaries of stability region with
respect to density fluctuations. Section \ref{Sec-Numerical}
contains the main results of the paper on numerical exploration of
the the phase diagram. After discussion of numerical
implementation of the model, we explore stability of the isolated
crossing configuration. In the next two subsections we consider
two different phase transitions between the tilted and crossing
chains and two types of phase diagrams which are realized in the
region of intermediate parameter $\alpha$.

\section{Energy functional\label{Sec-General}}

Our calculations are based on the Lawrence-Doniach free-energy functional in
the London approximation, which depends on the in-plane phases $\phi
_{n}(\mathbf{r})$ and vector-potential $\mathbf{A}(\mathbf{r})$
\begin{align}
F  &  =\sum_{n}\int d^{2}\mathbf{r}\left[  \frac{J}{2}\left(  \mathbf{\nabla
}_{\perp}\phi_{n}-\frac{2\pi}{\Phi_{0}}\mathbf{A}_{\perp}\right)  ^{2}\right.
\nonumber\\
&  \left.  +E_{J}\left(  1-\cos\left(  \phi_{n+1}-\phi_{n}-\frac{2\pi s}%
{\Phi_{0}}A_{z}\right)  \right)  \right] \nonumber\\
&  +\int d^{3}\mathbf{r}\frac{\mathbf{B}^{2}}{8\pi}, \label{LDen}%
\end{align}
where
\begin{equation}
J\equiv\frac{s\varepsilon_{0}}{\pi}\;\text{and }E_{J}\equiv\frac
{\varepsilon_{0}}{\pi s\gamma^{2}}%
\end{equation}
are the phase stiffness and the Josephson coupling energy with $\varepsilon
_{0}\equiv\Phi_{0}^{2}/\left(  4\pi\lambda\right)  ^{2}$, $\lambda
\equiv\lambda_{ab}$ and $\lambda_{c}$ are the components of the London
penetration depth, $\gamma=\lambda_{c}/\lambda_{ab}$ is the anisotropy factor,
and $s$ is the interlayer periodicity. The ratio of the two energy scales
determines the most important length scale of the problem, the Josephson
length, $\lambda_{J}=\gamma s=\sqrt{J/E_{J}}$. We use the London gauge,
$\mathrm{div}\mathbf{A}=0$. We mainly address the situation when magnetic
$\mathbf{B}$ inside the superconductor is fixed. The $c$ component of the
field determines the concentration of the pancake vortices $n_{v}\equiv
B_{z}/\Phi_{0}$ inside one layer. The in-plane phases $\phi_{n}$ have
singularities at the positions of pancake vortices $\mathbf{R}_{n,i}$ inside
the layers,
\[
\left[  \mathbf{\nabla}\times\mathbf{\nabla}\phi_{n}\right]  _{z}=2\pi\sum
_{i}\delta\left(  \mathbf{r}-\mathbf{R}_{n,i}\right)  .
\]
Logarithmic divergencies in the vicinity of pancake-vortex cores have to be
cut at the coherence length $\xi_{ab}$. A useful approach for superconductors
with weak Josephson coupling is to split the phase and vector-potential into
the vortex and regular contributions, $\phi_{n}=\phi_{vn}+\phi_{rn}$ and
$\mathbf{A}=\mathbf{A}_{v}+\mathbf{A}_{r}$. The vortex contributions minimize
the energy for fixed positions of pancake vortices at $E_{J}=0$ and give
magnetic interaction energy for the pancake vortices. One can express this
part of energy via the vortex coordinates $\mathbf{R}_{n,i}$. In general, the
regular contributions may include phases and vector-potentials of the
Josephson vortices. The total energy naturally splits into the regular part
$F_{r}$, the energy of magnetic interactions between pancakes $F_{M}$, and the
Josephson energy $F_{J}$, which couples the regular and vortex degrees of
freedom,
\begin{equation}
F=F_{r}+F_{M}+F_{J} \label{splitLDen}%
\end{equation}
with
\begin{widetext}
\begin{align}
F_{r}\left[  \phi_{rn},\mathbf{A}_{r}\right]   &  =\sum_{n}\int d^{2}%
\mathbf{r}\frac{J}{2}\left(  \mathbf{\nabla}\phi_{rn}-\frac{2\pi}{\Phi_{0}%
}\mathbf{A}_{r\perp}\right)  ^{2}+\int d^{3}\mathbf{r}\frac{\mathbf{B}_{r}%
^{2}}{8\pi},\label{Fr}\\
F_{M}\left[  \mathbf{R}_{n,i}\right]   &  =\frac{1}{2}\sum_{n,m,i,j}%
U_{M}(\mathbf{R}_{n,i}-\mathbf{R}_{m,j},n-m),\label{Fv}\\
F_{J}\left[  \phi_{rn},\mathbf{A}_{r},\mathbf{R}_{n,i}\right]   &  =\sum
_{n}\int d^{2}\mathbf{r}E_{J}\left(  1-\cos\left(  \nabla_{n}\phi_{n}%
-\frac{2\pi s}{\Phi_{0}}A_{z}\right)  \right)  , \label{FJ}%
\end{align}
where the discrete gradient $\nabla_{n}\phi_{n}$ is defined as
$\nabla_{n}\phi _{n}\equiv\phi_{n+1}-\phi_{n}$, and
$U_{M}(\mathbf{R},n)$ is the magnetic
interaction between pancakes\cite{pancakes}%
\begin{align}
U_{M}(\mathbf{R},n)  &  \approx2\pi J\left[  \ln\frac{L}{R}\left[  \delta
_{n}-\frac{s}{2\lambda}\exp\left(  -\frac{s|n|}{\lambda}\right)  \right]
+\frac{s}{4\lambda}u\left(  \frac{r}{\lambda},\frac{s|n|}{\lambda}\right)
\right] \label{MagInter}\\
u\left(  r,z\right)   &  \equiv\exp(-z)E_{1}\left(  r-z\right)  +\exp
(z)E_{1}\left(  r+z\right)  ,\nonumber
\end{align}
\end{widetext}
where $E_{1}(u)=\int_{u}^{\infty}\left(  \exp(-v)/v\right)  dv$ is the
integral exponent ($E_{1}(u)\approx-\gamma_{E}-\ln u+u\ $at$\;u\ll1$ with
$\gamma_{E}\approx0.5772$ being the Euler constant), $r\equiv\sqrt
{R^{2}+(ns)^{2}}$, and $L$ is a cutoff length. The regular and vortex degrees
of freedoms are coupled only via the Josephson energy in which $\phi_{n}$ is
the total phase composed of vortex and regular contributions.

The discrete layer structure has strongest influence on the cores
of tilted and Josephson vortices. Interaction contributions to the
total energy usually can be computed within continuous
approximation which describes the layered superconductor as a
three-dimensional anisotropic material. This approximation
amounts to replacement of summation in the layer index $n$ in Eqs.\ (\ref{Fr}%
), (\ref{Fv}), and (\ref{FJ}) with integration in the continuous
variable $z=ns$ and expansion of cosine in Eq.\ (\ref{FJ}). In the
continuous approximation one can derive a very useful general
result for the energy (\ref{LDen}) (see Ref.\
\onlinecite{SudboBrPRB91})
\begin{equation}
F=\frac{\Phi_{0}^{2}}{8\pi}\int\frac{d^{3}\mathbf{k}}{(2\pi)^{3}}%
\frac{(1+\lambda_{c}^{2}k^{2})\left\vert S_{z}\right\vert ^{2}+(1+\lambda
^{2}k^{2})\left\vert S_{\parallel}\right\vert ^{2}}{(1+\lambda^{2}%
k^{2})\left(  1+\lambda^{2}k_{z}^{2}+\lambda_{c}^{2}k_{\parallel}^{2}\right)
} \label{LondonVortEner}%
\end{equation}
in terms of vorticity $\mathbf{S}(\mathbf{r})$ of parametrically defined
vortex lines $\mathbf{R}_{i}(X)$
\[
\mathbf{S}(\mathbf{r})=\sum_{j}\int dX\frac{d\mathbf{R}_{i}}{dX}%
\delta(\mathbf{r}-\mathbf{R}_{i}(X))
\]
whose Fourier transform is
\[\mathbf{S}(\mathbf{k})=\sum_{j}\int
dX\frac {d\mathbf{R}_{j}}{dX}\exp(\imath\mathbf{kR}_{j}(X)).\] As
we will use this formula only for evaluation of interaction
energies between vortex lines, we have to subtract from it the
logarithmically diverging single-vortex terms.

In this paper we focus on the structure of an isolated vortex
chain with period $a$ in $x$ direction and period $c=Ns$ in $z$
direction corresponding to the tilting angle $\theta$ of vortices
with respect to the c axis with $\nu\equiv \tan\theta=a/c$ (see
Fig.\ \ref{Fig-CrossTiltChain}). The vertical period is fixed by
the in-plane field $B_{x}$,
$c\approx\sqrt{2\Phi_{0}/(\sqrt{3}\gamma B_{x})}$. For BSCCO
$\gamma\sim500$ and this period is approximately equal to $20$
layers at $B_{x}\approx50$G. We consider the case $c\ll\lambda$
and in-plane distances much smaller than $\lambda_{c}$. For this
particular problem the general energy given by Eqs.\
(\ref{splitLDen}), (\ref{Fr}), (\ref{Fv}), and (\ref{FJ}) can be
significantly simplified using several approximations: (i) we can
neglect screening of regular phase and $z$-axis vector-potential;
(ii)we consider only one-dimensional displacements of pancake rows
along the chain, $\mathbf{R}_{n,i}=(ai+u_{n},0,ns)$; (iii)we
subtract the energy of straight pancake stacks,
$(B_{x}/\Phi_{0})\varepsilon _{\mathrm{PS}}$ with $\varepsilon
_{\mathrm{PS}}\approx \varepsilon _{0}(\ln \kappa +0.5)$, allowing
us to eliminate logarithmically diverging pancake-core
contributions; (iv)we drop the trivial magnetic energy term
$B_{x}^{2}/8\pi$ which plays no role in selection between
different chain phases.
We will use the chain energy per unit area, $E\equiv c_y(F/V-B_{x}^{2}/(8\pi)-(B_{z}%
/\Phi_{0})\varepsilon_{\mathrm{PS}})$ with $V$ being the total system volume.
With the above assumptions this energy can be represented as
\begin{widetext}
\begin{align}
E & =\frac{1}{sN}\sum_{n=1}^{N}\int\limits_{0}^{a}\frac{dx}{a}\int
\limits_{-c_{y}/2}^{c_{y}/2}dy\left[  \frac{J}{2}\left( \nabla\phi
_{r,n}\right)  ^{2}+E_{J}\left(  1-\cos\left( \nabla_{n}\left(
\phi _{r,n}+\phi_{v,n}\right)  -\frac{2\pi
s}{\Phi_{0}}B_{x}y\right)  \right)
\right]  \label{ChainEn}\\
&  +\frac{1}{2sL_z}\sum_{n\neq m}U_{Mr}(u_{n}-u_{m},n-m)\nonumber
\end{align}
where $\phi_{v,n}\left(  \mathbf{r}\right)  $ is the vortex phase
variation induced by displacement of pancake rows, $u_{n}$, from
the ideally aligned
positions%
\begin{equation}
\varphi_{v,n}(x,y;u_{n})=\arctan\frac{\tan\left(
\pi(x+u_{n})/a\right) }{\tanh\left(  \pi y/a\right) }\nonumber
-\arctan\frac{\tan\left(  \pi x/a\right)
}{\tanh\left(  \pi y/a\right)  },\label{VortRowPhase}%
\end{equation}
\end{widetext}
$U_{Mr}(u_{n}-u_{m},n-m)$ is the interaction energy between the
pancake rows per unit length, computed with respect to straight
stacks
\begin{equation}
U_{Mr}(u,n)\!\equiv\!\frac{1}{a}\sum_{m}\left[
U_{M}(ma\!+\!u,n)\!-\!U_{M}(ma,n)\right], \label{UMr}
\end{equation}
$c_{y}=\Phi_{0}/cB_{x}\gg c$ is the in-plane distance between
chains, and $L_z$ is the total system length in $z$ direction.

The energy (\ref{ChainEn}) contains a long-range suppression of
the Josephson energy accumulated from distances $c\ll y\ll c_{y}$,
that is identical in all chain phases and it is convenient to
separate this term too. The averaged $z$-axis phase gradient
induced by the Josephson vortex lattice is given by
\[
\overline{\nabla_{n}\phi_{n}}-\frac{2\pi s}{\Phi_{0}}B_{x}y=\frac{\pi}%
{N}-\frac{2\pi y}{Nc_{y}}%
\]
Evaluating integral%
\[
\int_{-c_{y}/2}^{c_{y}/2}dy\left(  1-\cos\left(  \nabla_{n}\bar{\phi}%
_{n}-\frac{2\pi s}{\Phi_{0}}B_{x}y\right)  \right)  \approx\frac{\pi^{2}c_{y}%
}{6N^{2}},
\]
we obtain the long-range Josephson energy, $E_{\text{\textsc{J-LR}}%
}$,
\begin{equation}
E_{\text{\textsc{J-LR}}}=E_{J}\frac{\pi^{2}c_{y}}{6sN^2}%
\label{Long-RangeEn}%
\end{equation}
We will define the local energy, $E_{loc}$, sensitive to
the chain structure as%
\begin{equation}
E_{loc}\equiv E-E_{\text{\textsc{J-LR}}%
}.\label{Local-En}%
\end{equation}
This part of energy weakly depends on $c_{y}$ and does not diverge
for $c_{y}\rightarrow\infty$. The result (\ref{Long-RangeEn}) is
valid for chains separated by distance $c_y$ smaller than
$\lambda_c$. Similar calculation can be made for isolated chain
separated from other chains by distance larger than $\lambda_c$.
In this situation the integral over $y$ converges on distance
$\lambda_c$ instead of $c_y$ leading to result
$E_{\text{\textsc{J-LR}}}=E_{J}\pi \lambda_c/sN^2$. We will use
this result in the analytical calculations of the isolated chain
energy.

In calculation of magnetic coupling energy one has to take into
account periodic conditions for pancake displacements,
$u_{n+N}=u_{n}$. In addition, if one selects $z$ axis origin at
the center of the Josephson vortex then symmetry also requires
$u_{-n}=-u_{n}$. Ground state of the vortex chain is determined by
the minimum of energy (\ref{ChainEn}) with respect to pancake
displacements and regular phase distribution. Two simple limiting
cases in Fig.\ \ref{Fig-CrossTiltChain} correspond to (i)
$u_{n}\ll a$ for the crossing-chain configuration and (ii)
$u_{n}=-a(1-(n-1/2)/N_{l})/2$ for the tilted-chain configuration.

\section{Analytical calculations of chain energies \label{Sec-Analyt}}

In this section we compute energy of an isolated vortex chain with
period $a$ in $x$ direction and period $c=Ns\ll\lambda$ in $z$
direction. In general, there are two approaches to compute
energies of vortex configurations. Using distribution of phase and
vector potential, the total energy can be obtained directly from
the Lawrence-Doniach functional by integration of the local
energy. This approach is always used in numerical computations.
Analytically, it is more convenient to calculate the total energy
by summing up energy of isolated vortices and vortex interactions.
Analytic estimates for energy contributions are possible in two
limiting cases of weakly deformed crossing chain and chain
consisting of tilted vortices (see Fig.\
\ref{Fig-CrossTiltChain}). Comparison of these energies gives an
approximate range of parameters where one of these competing
configurations is energetically preferable. In contrast to the
numerical part, we consider isolated chain separated from other
chains by distance $c_y\gg \lambda_c$. For comparison with
numerical calculations, it will be necessary to extract the local
part of energy which is not sensitive to the long-range behavior.

\subsection{Crossing chains}

Energy of crossing-lattices chain per unit area is given by sum of
pancake-stack ($E_{PS}$), Josephson-vortex ($E_{JV}$) and crossing energies
($E_{\times}$) terms,%
\begin{equation}
E_{CL}=E_{PS}+E_{JV}+E_{\times}\text{.} \label{EnCrossChain}%
\end{equation}
Both pancake and JV terms are composed of single-vortex and interaction
contribution, $E_{PS}=E_{PS}^{s}+E_{PS}^{i}$, $E_{JV}=E_{JV}^{s}+E_{JV}^{i}$.

We start with evaluating pancake-stack energies. Contribution from energies of
individual stacks to the energy per unit length is given by%
\begin{equation}
E_{PS}^{s}=\frac{\varepsilon_{0}}{a}\left(  \ln\frac{\lambda}{\xi}%
+C_{v}\right)  , \label{En_PS_s}%
\end{equation}
where $C_{v}\approx0.497$ within the Ginzburg-Landau theory. Using
vorticity of the pancake-vortex chain, $S_{z}(\mathbf{r})
=\sum_{j}\delta(y)\delta(x-ja)$ corresponding to
$S_{z}(\mathbf{k}) =\delta(k_{z})\sum_{j}\exp\left(
ik_{x}ja\right)  $, we derive from general formula
(\ref{LondonVortEner}) the
pancake-stacks interaction energy%
\begin{equation}
E_{PS}^{i}=\frac{\Phi_{0}^{2}}{8\pi a}\int\frac{dk_{x}dk_{y}}{(2\pi)^{2}}%
\sum_{j\neq0}\frac{\exp\left(  ik_{x}ja\right)  }{1+\lambda^{2}k^{2}}%
\end{equation}
Using relation $\sum_{j}\exp\left(  ik_{x}ja\right)  )=\frac{1}{a}\sum
_{m}\delta(k_{x}-\frac{2\pi m}{a})$ and integrating over $k_{y}$ we obtain%
\begin{align*}
E_{PS}^{i}&=\frac{\Phi_{0}^{2}}{16\pi a^{2}\lambda}\sum_{m=-\infty
}^{\infty}\frac{1}{\sqrt{1+(2\pi\lambda
m/a)^{2}}}\\&-\frac{\Phi_{0}^{2}}{16\pi
a\lambda}\int\frac{dk}{2\pi}\frac{1}{\sqrt{1+\lambda^{2}k^{2}}}
\end{align*} For comparison with the energy of tilted chain, it
will be convenient to represent this energy in the form
\begin{equation}
E_{PS}^{i}=\frac{\varepsilon_{0}}{a}\left(
\frac{\pi\lambda}{a}-\ln\frac{4\pi \lambda}{a}+\gamma_{E}-U\left(
\frac{a}{2\pi\lambda}\right)  \right) \label{En_PS_i}
\end{equation}
with
\[
U\left(  x\right) \! =\!\sum_{m=1}^{\infty}\left(
\frac{1}{m}\!-\!\frac{1} {\sqrt{m^{2}+x^{2}}}\right) \! =\!
\genfrac{\{}{.}{0pt}{}{\zeta(3)x^{2}/2,\;x\lesssim0.5}{\frac{1}{2x}-\ln
\frac{2}{x}+\gamma_{E},\;x\gtrsim1},
\]
and $\zeta(3)\equiv\sum_{m=1}^{\infty}\frac{1}{m^{3}}\approx1.202$

Single-vortex JV energy is given by \cite{ClemJosLat,Kinkwalls}%
\[
E_{JV}^{s}=\frac{\varepsilon_{0}}{\gamma c}\left(  \ln\frac{\lambda}%
{s}+1.54\right)
\]
The JV interaction energy can be evaluated in the same way as the
pancake-stack interaction energy (\ref{En_PS_i}) and in the limit
$c\ll\lambda$ it is
given by%
\[
E_{JV}^{i}\approx\frac{\varepsilon_{0}}{\gamma c}\left(  \frac{\pi\lambda}%
{c}-\ln\frac{4\pi\lambda}{c}+\gamma_{E}\right)  .
\]
Therefore, the total energy of the JV array, $E_{JV}=E_{JV}^{s}+E_{JV}^{i}$,
can be written as%
\begin{equation}
E_{JV}=\frac{\varepsilon_{0}}{\gamma c}\left(  \frac{\pi\lambda}{c}+\ln
\frac{c}{s}+C_{JV}\right)  \label{En_JVArray}%
\end{equation}
with $C_{JV}=1.54-\ln4\pi+\gamma_{E}\approx-0.41$. The fist term
in this formula is the long-range Josephson energy,
$E_{\mathrm{J-LR}}$. This term is identical in all chain phases
and does not determine selection between them.  This formula is
valid for an isolated chain and it is different from the
long-range Josephson energy of the dense JV lattice,
$c_y<\lambda_c$, given by Eq.\ (\ref{Long-RangeEn}). The
difference amounts to a simple replacement $c_y\rightarrow
\pi\lambda_c$.  For comparison with simulations, we will use only
the local part of energy which is obtained from the total energy
by subtracting $E_{\mathrm{J-LR}}$.

Using estimate for the crossing energy of Josephson vortex and
pancake stack for $\alpha=\lambda/\gamma s\lesssim 0.4$
\cite{CrossLatPRL99,JVpancPRB03}
\begin{equation}
\epsilon_{\times}=-\frac{8\alpha^{2}s\varepsilon_{0}}{\ln(3.5/\alpha)},
\label{eps_x}%
\end{equation}
we obtain the contribution from the crossings into the energy per unit area%
\begin{equation}
E_{\times}=\frac{\epsilon_{\times}}{ac}=-\frac{8\alpha^{2}\varepsilon_{0}}%
{\ln(3.5/\alpha)aN}. \label{Ex}%
\end{equation}
Finally, combining results (\ref{En_PS_s}), (\ref{En_PS_i}), (\ref{En_JVArray}%
) and (\ref{Ex}) we obtain the total energy of crossing chain%
\begin{widetext}
\begin{equation}
E_{CL}=E_{PS}^{s}+\frac{\varepsilon_{0}}{a}\left(  \frac{\pi\lambda}{a}%
-\ln\frac{4\pi\lambda}{a}+\gamma_{E}-U\left(  \frac{a}{2\pi\lambda}\right)
+\frac{\pi\nu\lambda}{\gamma c}+\frac{\nu}{\gamma}\left(  \ln N+C_{JV}\right)
-\frac{8\alpha^{2}}{\ln(3.5/\alpha)N}\right)  . \label{EnCrossChain_res}%
\end{equation}
\end{widetext}
Subtracting the pancake-stack and long-range Josephson energies,
we obtain local energy,
$E_{CL}^{loc}=E_{CL}-E_{PS}-\varepsilon_{0}\pi\lambda/\gamma
c^{2}$,
\begin{equation}
E_{CL}^{loc}=\frac{\varepsilon_{0}}{a}\left(  \frac{\nu}{\gamma}\left(  \ln
N+C_{JV}\right)  -\frac{8\alpha^{2}}{\ln(3.5/\alpha)N}\right)  ,
\label{EnCrossChain_loc}%
\end{equation}
which we will use for comparison with numerical simulations.

\subsection{Tilted chains}

Energy of the tilted chain per unit area also can be decomposed into the
single-vortex and interaction contributions%
\[
E_{TV}=E_{TV}^{s}+E_{TV}^{i}.
\]
The first term can be estimated analytically only in two limiting
cases $\tan\theta\ll\gamma$ and $\tan\theta>\gamma$, where the
second case corresponds to kinked lines. For the interaction term
we derive a general formula valid in both limits.

\subsubsection{Energy of a single tilted line}

\paragraph{Region $\nu=\tan\theta<\gamma$}

Energy difference between tilted and straight pancake stacks, $\varepsilon
_{TV}-\varepsilon_{PS}$, determines tilt stiffness of pancake stack and
contains magnetic and Josephson contribution, $\varepsilon_{M}$ and
$\varepsilon_{J}$. Magnetic part has been calculated by Clem \cite{ClemPRB91}%
\[
\varepsilon_{M}=\varepsilon_{0}\ln\frac{\sqrt{1+\nu^{2}}+1}{2}.
\]
The Josephson contribution to the tilt energy appears due to
suppression of the Josephson interlayer coupling by mismatched
pancakes. It can be evaluated as
\[
\varepsilon_{J}=\varepsilon_{0}\frac{\nu^{2}}{2\gamma^{2}}\left(
\ln \frac{\lambda_{c}}{s\nu}+C_{J}\right) .
\]
Numerical constant $C_{J}$ can be computed exactly by matching logarithmic
contributions coming from small and large distances \cite{DodgsonAEK} which
gives $C_{J}=3\ln2-\gamma_{E}$. Therefore, the single-vortex energy of the
tilted chain is given by%
\begin{equation}
E_{TV}^{s}\!=\!E_{PS}^{s}\!+\!\frac{\varepsilon_{0}}{a}\ln\frac{\sqrt{1+\nu^{2}}+1}%
{2}\!+\!\frac{\varepsilon_{0}}{a}\frac{\nu^{2}}{2\gamma^{2}}\left(
\ln
\frac{8\lambda_{c}}{s\nu}\!-\!\gamma_{E}\right)  \label{Etlt-s}%
\end{equation}
Note that at not very large tilt angles, $\nu \lesssim 1$, the
Josephson tilt energy is roughly $\gamma^{2}$ smaller than the
magnetic tilt energy. Two terms become comparable at very large
tilt angles, $\nu \sim \gamma$. In several papers
\cite{SudboBrPRB91,TiltedChains} the energy of a tilted line in
\emph{anisotropic three-dimensional} superconductor has been
calculated. This calculation is based on the London energy
expression (\ref{LondonVortEner}) employing elliptic cut off in
$k$-integration $k_\Vert^2+ k_z^2/\gamma^2<\xi_{ab}^{-2}$ to treat
the vortex core. Strictly speaking, this approximation does not
describe layered superconductors in which $\xi_{c} < s$.
Nevertheless, in the region $\nu<\gamma$ it leads to the result
similar to Eq.\ (\ref{Etlt-s}) with somewhat different magnetic
part and different expression under logarithm in the Josephson
part.

\paragraph{Region $\nu=\tan\theta>\gamma$ (kinked lines).}

In the region $\tan\theta>\gamma$ the vortex lines have the kink
structure, i.e., they are composed of kinks separated by JV
pieces. Energy of such line per unit length is given by
\begin{equation}
\varepsilon_{\mathrm{kl}}=\varepsilon_{JV}+u_{k}/L+\varepsilon_{\mathrm{ki}}
\label{En_kink_line}%
\end{equation}
where $\varepsilon_{JV}=(\varepsilon_{0}/\gamma)\left(
\ln(\lambda /s)+1.54\right)  $ is the energy of Josephson vortex,
\begin{equation}
u_{k}=s\varepsilon _{0}\left(  \ln(\gamma s/\xi)+C_{k}\right)
\label{kink_Ener}
\end{equation}
is the kink energy \cite{BLK,Kinkwalls}, $L=s\tan\theta$ is the
separation between kinks, and $\varepsilon_{\mathrm{ki}}$ is the
kink interaction energy. Numerical constant $C_{k}$ in the kink
energy has been estimated as $-0.17$ in Ref.
\onlinecite{Kinkwalls}. However more accurate numerical
calculations of this paper give somewhat smaller value
$C_{k}\approx-0.31$.

The interaction energy between neighboring kinks in the kinked line decays as
$1/L^{2}$ up to $L<\lambda_{c}$ leading to relatively large interaction
contribution to the total energy. The kink interaction energy is computed in
details in the Appendix \ref{App:KinkInter}. In the region $L\ll\lambda_{c}$
the dominant interaction term in the energy of single line is given by
\begin{equation}
\varepsilon_{\mathrm{ki}}=\frac{\gamma s^{2}\varepsilon_{0}}{2L^{2}}\left(
\ln\left(  \frac{\lambda_{c}}{L}\right)  -\frac{3}{2}\right)
\label{En_ki_result}%
\end{equation}

Combining all contributions, we derive the following result for
the single-vortex energy of a tilted chain,
$E_{TV}^{s}=\varepsilon_{\mathrm{kl}}/c$, in the limit
$\nu >\gamma$%
\begin{align}
E_{TV}^{s}&=E_{PS}^{s}+E_{JV}^{s}\nonumber\\&
+\frac{\varepsilon_{0}}{a}\left( \ln\frac
{1}{\alpha}+C_{kv}+\frac{\gamma}{2\nu}\left(  \ln\left(  \frac{\lambda_{c}N}%
{a}\right)  -\frac{3}{2}\right)  \right)  \label{E_TVsingle_large}%
\end{align}
with $C_{kv}\equiv C_{k}-C_{v}\approx-0.81$. Note that the values
of the numerical constants $C_{v}$ and $C_{k}$ depend on the core
structure at small distances $r\sim\xi$ from its center, which
evolves with temperature decrease. However, as this structure is
exactly the same for the pancake vortex and kink, the difference
$C_{k}-C_{v}$ is not sensitive to behavior at small distances and
remains the same down to low temperatures. Criterion $\ln\left(
1/\alpha_{c}\right)  +C_{kv}=0$ (corresponding to
$\alpha_{c}\approx0.44$) separates the kink and pancake-stack
penetration regimes of a small c-axis field for large $N$
(somewhat larger value $\alpha_{c}\approx 0.5$ has been given in
Ref.\ \onlinecite{Kinkwalls}). This critical value increases with
decrease of $N$ due to the crossing energies in the crossing
chains. Penetration of the c-axis field in the presence on the
in-plane field is frequently referred to as a lock-in transition.

\subsubsection{Interaction energy of tilted vortices.}

The interaction energy between tilted lines is not influenced much
by the layered structure and it can be computed within the London
approximation. In  Appendix \ref{App:InterTilted} we derive a
general analytical formula for the interaction energy of tilted
chain, $E_{TV}^{i}$,
\begin{widetext}
\begin{equation}
E_{TV}^{i}\!    \approx\!\frac{\varepsilon_{0}}{a}\left[
\frac{\pi\nu \lambda}{\gamma
c}\!+\!\frac{\pi\lambda}{a}\!-\!\frac{\sqrt
{\nu^{2}+\gamma^{2}}}{\gamma}\left(  \!\ln\left(
\frac{4\pi\lambda\sqrt {\nu^{2}+\gamma^{2}}}{a}\right)
\!-\!\gamma_{E}\right)  \!+\!\ln\frac{\gamma+\sqrt{\nu^{2}+\gamma^{2}}}{1+\sqrt{1+\nu^{2}}%
}\!\right] \label{EnTiltInt}
\end{equation}
The first term in this formula represents again the long-range
Josephson energy and it is identical to the first term in the
energy of the Josephson chain (\ref{En_JVArray}). At large $N$ the
above formula also gives the interaction energy of a kinked line
for $\nu>\gamma$, because the kinked structure of tilted lines
starts to influence their interaction only at very large tilting
angle $\nu>N\gamma/2\pi$.\cite{note:KinkWalls} In the region
$\nu\ll\gamma$ Eq.\ (\ref{EnTiltInt}) simplifies as
\begin{equation}
E_{TV}^{i}  \approx\frac{\varepsilon_{0}}{a}\left[  \frac{\pi\nu^{2}%
\lambda}{\gamma a}+\frac{\pi\lambda}{a}-\ln\left(  \frac{2\pi
\lambda\left(  1+\sqrt{1+\nu^{2}}\right)  }{a}\right) +\gamma_{E}
-\frac{\nu^{2}}{2\gamma^{2}}\left( \!\ln\left( \frac{4\pi
\lambda_{c}}{a}\right) -\gamma_{E}+\frac{1}{2}\right) \right]
\label{EnTiltIntSmall}
\end{equation}
\end{widetext}
The last term represents change of the Josephson energy due to
misalignment of pancakes in different stacks. Combining this term
with the last term in Eq. (\ref{Etlt-s}), we obtain the total
Josephson energy loss of tilted chain due
to pancake misalignment%
\[
\delta E_{TV}=\frac{\varepsilon_{0}}{a}\frac{\nu^{2}}{2\gamma^{2}}\left(
\ln\frac{2c}{\pi s}-\frac{1}{2}\right)
\]
In the limit of kinked lines, $\nu\gg\gamma$, interaction energy reduces to
\begin{align}
E_{TV}^{i}\!&\approx\!\frac{\varepsilon_{0}}{a}\left[ \frac{\pi\nu^{2}%
\lambda}{\gamma a}\!+\!\frac{\pi\lambda}{a}\!-\!\frac{\nu}{\gamma
}\left(  \ln\left(  \frac{4\pi\lambda}{c}\right)
-\!\gamma_{E}\right) \right.\nonumber\\
&\left.-\frac{\gamma}{2\nu}\left(  \ln\left(
\frac{4\pi\lambda}{c}\right)
-1-\!\gamma_{E}\right)  \!\right]  \label{EnTiltIntLarge}%
\end{align}
Note again that this result is valid until $\nu<N\gamma/2\pi$,
where the kink structure of the tilted lines does not influence
much interaction between them. The limit of larger $\nu$
corresponds to the regime of ``kink walls'' described in Ref.\
\onlinecite{Kinkwalls}.

\subsubsection{Total energy of tilted chains}

Combining the interaction energy (\ref{EnTiltIntSmall}) with the
energy of individual stacks (\ref{Etlt-s}), we obtain the total
tilted-chain energy in the limits $c=Ns\ll\lambda$, $\nu\ll\gamma$
\begin{align}
E_{TV}&\approx E_{PS}^{s}+\frac{\varepsilon_{0}}{a}\left[
\frac{\pi \nu \lambda }{\gamma c}+\frac{\pi\lambda}{a}-\ln\left(
\frac{4\pi\lambda}{a}\right)
+\gamma_{E}\right.\nonumber\\
&\left.+\frac{\nu^{2}}{2\gamma^{2}}\left(  \ln N-0.95\right)
\right] .
\label{E_TVresult_sm}%
\end{align}
This gives the following result for the local part of energy, $E_{TV}%
^{loc}\equiv E_{TV}-E_{PS}-\varepsilon_{0}\pi\lambda/\gamma c^{2}$,%
\begin{equation}
E_{TV}^{loc}=\frac{\varepsilon_{0}}{a}\left[  U\left(  \frac{a}{2\pi\lambda
}\right)  +\frac{\nu^{2}}{2\gamma^{2}}\left(  \ln N-0.95\right)  \right]
\label{E_TVsm_local}%
\end{equation}
In this formula the first term represents the loss of the magnetic
coupling energy in the tilted chain and the second term represents
the Josephson energy loss.

In the region of kinked lines $\nu\gg\gamma$ (but $\nu\ll
N\gamma/2\pi$), the total chain energy is obtained by combining
Eqs.\ (\ref{E_TVsingle_large}) and
(\ref{EnTiltIntLarge}) giving%
\begin{align}
E_{TV}&\approx
E_{PS}^{s}+E_{JV}\nonumber\\
&+\frac{\varepsilon_{0}}{a}\left[ \ln
\frac{\gamma s}{\lambda}+C_{kv}+\frac{\gamma}{2\nu}\left(
\ln\left( \frac{\gamma N}{\nu}\right)  +C_{\mathrm{ki}}\right)
\!\right]
\label{E_TVresult_lrg}%
\end{align}
with $C_{\mathrm{ki}}=-\ln4\pi-1/2+\gamma_{E}\approx-2.454$.

\section{Location of transitional regions in the phase space
\label{Sec-TransRegion}}

To find out whether the crossing or tilted chain is realized for
given values of the parameters $a$, $c$, and $\alpha$, we have to
compare the energies of these states. Naively, one may think that
intersection of the energy curves for the two states would
correspond to a first-order phase transition between these states.
However, as we will see from numerical simulations, in the region
$\nu=\tan\theta<\gamma$ another scenario is realized. Typically,
strongly deformed intermediate chain configurations develop in the
transitional region providing a smooth transition between the two
limiting configurations. Therefore, a simple energy comparison
gives only an approximate location of the transitional region
separating the two configurations.

In the region $\nu/\gamma=a/N\lambda_{J}\ll1$ comparison of
(\ref{EnCrossChain_res}) and (\ref{E_TVresult_sm}) gives the
following criterion for the transitional region
\begin{align}
&U\left(  \frac{a}{2\pi\lambda}\right)  -\frac{\nu}{\gamma}\left(
\ln N-0.41\right)  \nonumber\\
&+\frac{\nu^{2}}{2\gamma^{2}}\left(
\ln N-0.95\right)
+\frac{8\alpha^{2}}{\ln(3.5/\alpha)N}=0 \label{Trans_Crit_sm}%
\end{align}
One can observe that the main competition takes place between the
loss of the magnetic coupling energy for the tilted chain (first
term) and strong suppression of the Josephson energy by JVs for
the crossing chain (second term). Solution of this equation
provides the boundary which can be written in the reduced form
$a=\lambda_{J}N\ f(N,\alpha)$ with $f(N,\alpha)\ll1$, i.e., the
boundary shape in the plane $a/\lambda_J$-$N$ depends only on the
parameter $\alpha$.

In region $1<\nu/\gamma<N/2\pi$ and $a>2\pi\lambda$ comparison of the energies
(\ref{EnCrossChain_res}) and (\ref{E_TVresult_lrg}) gives%
\begin{equation}
\ln\frac{1}{\alpha}\!+\!C_{kv}\!+\!\frac{\gamma}{2\nu}\left(
\ln\left( \frac{\gamma N}{\nu}\right)\!+\!C_{\mathrm{ki}}\right)
+\frac{8\alpha^{2}}{\ln(3.5/\alpha
)N}\!=\!0\! \label{Tran_Crit_Lrg}%
\end{equation}
This equation has a solution only in the kink penetration regime,
$\ln\left( 1/\alpha\right) +C_{kv}<0$, near the transition between
the two penetration regimes $\left\vert \ln\left(  1/\alpha\right)
+C_{kv}\right\vert \ll1$ where the kink energy is only slightly
smaller than the energy per pancake of a straight pancake-vortex
stack. In contrast to Eq.\ (\ref{Trans_Crit_sm}), which gives only
an approximate location of the broad transitional region, this
equation indeed describes a very strong first-order phase
transition.

\section{Attraction between deformed pancake stacks and tilted vortices.
Maximum equilibrium separation \label{Sec-Attrac}}

A peculiar property of the crossing chain is an attractive
interaction between the deformed pancake stacks at large
distances.\cite{BuzdinPRL02} As a consequence, when the magnetic
field is tilted from the layer direction, the density of the
pancake stacks located on the Josephson vortices jumps from zero
to a finite value. This means the existence of a maximum
equilibrium separation $a_{m}$ between pancake stacks, i.e.,
chains with $a>a_{m}$ are not realized in equilibrium. Note that
the tilted vortices also attract each other within some range of
angles and distances \cite{TiltedChains} meaning that tilted
chains also have this property in some range of parameters.

A simple analytical formula for the attraction energy between the
deformed pancake stacks can be derived for very anisotropic
superconductors $\lambda_{J}\gg\lambda$ in the range $\lambda\ll
R\ll\lambda_{J}$.\cite{BuzdinPRL02} In this limit short-range
pancake displacements $u_{n}$ from the aligned positions in the
two neighboring stacks produce a dipole-like contribution to the
interaction energy per unit length between these stacks
\[
\delta U_{i}(R)=-\frac{2\varepsilon_{0}}{NR^{2}}\sum_{n=1}^{N}u_{n}^{2}%
=-\frac{2\varepsilon_{0}\left\langle u^{2}\right\rangle }{R^{2}}.
\]
This term has to be combined with the usual repulsive interaction
between straight stacks
$U_{i0}(R)=2\varepsilon_{0}\mathrm{K}_{0}(R/\lambda
)\approx2\varepsilon_{0}\sqrt{\pi\lambda/2R}\exp\left(
-R/\lambda\right)  $. Minimum of the total interaction energy,
$U_{i0}(R)+\delta U_{i}(R)$, gives an estimate for the maximum
equilibrium separation $a_{m}$ \cite{BuzdinPRL02},
$a_{m}=\lambda\ln(C\lambda^{2}/\left\langle u^{2}\right\rangle )$
and this result is valid until $a_{m}<\lambda_{J}$. Because in
BSCCO $\lambda_{J}$ is only $2-3$ times larger than $\lambda$,
this simple formula is not practical for this compound. We will
see that $a_{m}$ in BSCCO is usually larger than $\lambda_{J}$.

One can obtain a useful general recipe for determination of the
maximum equilibrium separation between the pancake stacks directly
from the chain energy per unit area $E$, without splitting it into
the single-vortex and interaction parts. We consider situation
when the in-plane component of the magnetic field is much larger
than the lower critical field in this direction so that the
in-plane magnetic induction practically coincides with the
in-plane external magnetic field. The $c$ component of the
external field, $H_{z}$, determines the effective chemical
potential $\mu$ for density of pancake stacks,
$\mu=\Phi_{0}H_{z}/(4\pi )$. With fixed chemical potential, the
chain thermodynamic potential per unit area depends on pancake
linear density $n$ as
\begin{equation}
G(n)=E(n)-\mu n \label{En-mun}.
\end{equation}
The equilibrium density $n_{\mu}$ is given by the minimum of this energy%
\[
\frac{dE(n)}{dn}=\mu.
\]
Substituting this relation back to Eq. (\ref{En-mun}) we obtain%
\[
G(n_{\mu})=E(n_{\mu})-E^{\prime}(n_{\mu})n_{\mu}.
\]
If we represent the energy as a function of the stack separation $a$ rather
than the density, the last result can be rewritten in a more compact way%
\[
G(a)=\frac{d}{da}\left(  aE(a)\right)
\]
The separation $a$ in this formula corresponds an equilibrium
separation only if $G(a)$ is smaller than the energy of the
Josephson vortex lattice $G(\infty)\equiv E(\infty)=E_{JV}$.
Therefore the maximum equilibrium separation $a_{m}$ is given by
the condition
\begin{equation}
\frac{dU(a)}{da}=0 \label{Condition_a_m}%
\end{equation}
where the quantity $U(a)=a\left(  E(a)-E(\infty)\right)  $
represents the pancake part of energy per one stack and the
condition (\ref{Condition_a_m}) implies that $a_{m}$ is determined
by the minimum of this energy. When the main contribution to the
total interaction energy is coming from the nearest-neighbor
interaction, $a_{m}$ coincides with the position of the minimum in
the pair interaction potential.

In principle, nonequilibrium structures with $a>a_{m}$ can be
prepared by applying external stretching forces at the chain
edges. However, the chain can be stretched only up to a certain
critical value of separation. Above this value the system becomes
unstable with respect to density fluctuations leading to formation
of high-density chain clusters. The stability criterion of the
chain is given by
\begin{equation}
\frac{d^{2}E(n)}{dn^{2}}>0.
\end{equation}
As
\[
\frac{d^{2}E(n)}{dn^{2}}=a^{3}\frac{d^{2}}{da^{2}}\left[  a\left(
E(a)-E(\infty)\right)  \right],
\]
the stability criterion can also be written as%
\begin{equation}
\frac{d^{2}U(a)}{da^{2}}>0. \label{StabCrit}
\end{equation}
Therefore, in the dependence of the pancake part of energy per
unit cell, $U$, on pancake separation $a$, the minimum gives the
maximum equilibrium separation and the inflection point
corresponds to the boundary of instability with respect to cluster
formation.

As an example, we apply the obtained general formula
(\ref{Condition_a_m}) to the tilted chain at $\nu<\gamma$. Using
Eqs.\ (\ref{E_TVresult_sm}) and (\ref{En_JVArray}), we obtain from
Eq.\ (\ref{Condition_a_m}) a cubic equation for $a_{m}$
\begin{equation}
\frac{\tilde{a}_{m}^{3}}{N^{2}}\left(  \ln N-0.95\right)  -\frac{\tilde{a}%
_{m}^{2}}{N}\left(  \ln N-0.41\right)  +\tilde{a}_{m}-\pi\alpha=0,
\label{a_m_tilted}%
\end{equation}
with $\tilde{a}_{m}\equiv a_{m}/\lambda_{J}$, which can be easily
solved numerically in general case. At small $\alpha$ and large
$N$ an approximate solution of this equation gives%
\[a_m\approx
\pi \lambda \left( 1+\frac{\pi \alpha }{N}\left( \ln N-0.41\right)
\right),
\]
i.e., at large $N$ $a_m$ approaches a remarkably  simple universal
value $\pi \lambda$.

\section{Numerical exploration of chain structures \label{Sec-Numerical}}

\subsection{Numerical implementation of the model}

Our purpose is to calculate the equilibrium distribution of the
regular phase $\phi_{r,n}(x,y)$ and pancake-row displacements on
the basis of the energy (\ref{ChainEn}). To facilitate numerical
calculations, we introduce reduced variables
\[
\tilde{y}=y/\lambda_{J},\ \tilde{x}=x/a,\ v_{n}=u_{n}/a
\]
and represent the row interaction energy (\ref{UMr}) in the reduced form%
\[
U_{Mr}(u,n)=\frac{\pi Ja}{\lambda^{2}}\tilde{\mathcal{V}}_{Mr}(\frac{u}%
{a},n),
\]
where $\tilde{\mathcal{V}}_{Mr}(v,n)=\mathcal{V}_{Mr}(v,n)-\mathcal{V}%
_{Mr}(0,n)$ and for $\mathcal{V}_{Mr}(v,n)$ we derive from Eq.\
(\ref{MagInter})
\begin{widetext}
\begin{equation}
\mathcal{V}_{Mr}(v,n)=\frac{s\lambda}{2a^{2}}\left[  2\exp\left(  -\frac
{s|n|}{\lambda}\right)  \ln\left[  2\sin(\pi|v|)\right]  +\sum_{m=-\infty
}^{\infty}u\left(  \frac{\left\vert v-m\right\vert }{\lambda/a},\frac
{s|n|}{\lambda}\right)  \right]  . \label{VMr_reduced}%
\end{equation}
The reduced energy per unit area, $\tilde{E}= E\lambda_J
/\varepsilon_{0}$, takes the form
\begin{equation}
\tilde{E}  \!=\!\frac{1}{\pi N}\sum_{n=1}^{N}\int_{0}^{\tilde{a}}\frac{d\tilde{x}%
}{\tilde{a}}\int_{-\tilde{c}_{y}/2}^{\tilde{c}_{y}/2}\!d\tilde{y}\left[
\frac{1}{2}\left(  \nabla\phi_{r,n}\right)
^{2}\!+\!1\!-\!\cos\left( \nabla _{n}\left(
\phi_{r,n}\!+\!\phi_{v,n}\right) \!-\!h\tilde{y}\right)
\right]\!+\!\frac{\tilde{a}}{2\alpha^{2}N_{tot}}\sum_{n\neq m}\tilde{\mathcal{V}%
}_{Mr}(v_{n}\!-\!v_{m},n\!-\!m)\label{ReducedEn}
\end{equation}
\end{widetext}
with $h\equiv2\pi B_{x}\gamma s^{2}/\Phi_{0}$ and $\tilde{a}=a/\lambda_{J}$.
The reduced local energy is defined as%
\begin{equation}
\tilde{E}_{loc}=\tilde{E}-\frac{\pi\tilde{c}_{y}}{6N^2}
\label{ReducedLocEn}%
\end{equation}
In particular, from Eqs.\ (\ref{EnCrossChain_loc}) and
(\ref{E_TVsm_local}) we obtain the following results for the
reduced local energies of two limiting configurations in the
limit $\tilde{a}/N=\nu/\gamma\ll 1$%
\begin{equation}
\tilde{E}_{loc}\!\approx
\genfrac{\{}{.}{0pt}{}{  \frac{\ln N+C_{JV}}{N}  \!-\!\frac{8\alpha^{2}%
}{N\tilde{a}\ln(3.5/\alpha)},\ \text{for crossing chain}}{\frac{1}%
{\tilde{a}}U\!\left(  \frac{\tilde{a}}{2\pi\alpha}\right)  \!+\!\frac{\tilde{a}%
}{2N^2}\left(  \ln N-0.95\right)  ,\ \text{for tilted chain.}}
\label{Reduced_Eloc_est}%
\end{equation}
We will also use the excess pancake part of energy defined as
\begin{align}
\delta E&\equiv E-E_{JV}\label{panc-ener}\\
&=(\lambda_{J}/\varepsilon_{0})(E-E^{s}_{PS}-E_{JV})\label{panc-ener1}
\end{align}%
with $E_{JV}$ being the JV lattice energy per unit area and the
excess pancake energy per stack (in units of $\varepsilon_{0}$),
$\tilde{U}\equiv\tilde{a}\delta\tilde{E}$.

The phase distribution $\phi_{r,n}$ and pancake displacements
$v_{n}$ minimizing the energy functional (\ref{ReducedEn}) obey
the following equations
\begin{widetext}
\begin{subequations}
\label{ReducEq}%
\begin{align}
&  \Delta\phi_{r,n}\!+\!\sin\left(  \nabla_{n}\left(  \phi_{r,n}%
\!+\!\phi_{v,n}\right)  \!-\!h\tilde{y}\right)  \!-\!\sin\left(  \nabla
_{n}\left(  \phi_{r,n-1}\!+\!\phi_{v,n-1}\right)  \!-\!h\tilde{y}\right)
=0\label{ReducEq1}\\
&  \nabla_{y}\phi_{rn}(v_{n}\tilde{a},0)+\frac{\tilde{a}}{2\alpha^{2}}%
\sum_{m=-\infty}^{\infty}\mathcal{F}_{Mr}(v_{n}-v_{m},n-m)=0, \label{ReducEq2}%
\end{align}
where
$\mathcal{F}_{Mr}(v,n)\equiv-\nabla_{v}\mathcal{V}_{Mr}(v,n)$ is
the
magnetic interaction force between the vortex rows,%
\end{subequations}
\begin{equation}
\mathcal{F}_{Mr}(v,n)=-\frac{\pi s\lambda}{a^{2}}\exp\left(  -\frac
{s|n|}{\lambda}\right)  \cot\pi v+\frac{s\lambda}{a^{2}}\sum_{m=-\infty
}^{\infty}\frac{1}{v-m}\exp\left(  -\frac{\sqrt{\left(  v-m\right)  ^{2}%
+s^{2}n^{2}}}{\lambda/a}\right)  . \label{FMr_reduced}%
\end{equation}
\end{widetext}
Note that both $\mathcal{V}_{Mr}(v,n)$ and $\mathcal{F}_{Mr}(v,n)$ are regular
at $v\rightarrow0$, because divergency in the first term is compensated by the
$m\!=\!0$ term in the sum.

In numerical calculations we assume for simplicity periodic
boundary condition in $y$ direction with period $c_{y}$.
Physically, this corresponds to rectangular arrangement of both
Josephson vortices and pancake stacks. Even though such
arrangement does not give the true ground state, it does not
influence much the structure of an individual chain, that is the
focus of this paper. Due to symmetry properties, it is sufficient
to find the phase distribution $\phi_{n}(x,y)$ within domain
$0<x<a$, $0<y<c_{y}/2$, $1\leq
n\leq N_{l}=N/2$ with the following boundary conditions for the total phase%
\begin{align*}
&\phi_{n}(a+0,y)    =\phi_{n}(+0,y),\ \phi_{n}(x,-0)=\pi-\phi_{n}(x,+0),\\
&\phi_{n}\left(  x,\frac{c_{y}}{2}+0\right)     =-\phi_{n}\left(
x,\frac{c_{y}}{2}-0\right)  +\frac{2\pi(n-1/2)}{N},\\
&\phi_{0}(x,y)   =-\phi_{1}(-x,y),\ \phi_{N_{l}+1}(x,y)=\pi-\phi_{N_{l}%
}(x,y).
\end{align*}
Pancake displacements $v_{n}$ have symmetry properties
$v_{n+N}=v_{n}$ and $v_{1-n}=-v_{n}$. This allows us to reduce
infinite $m$ summation in Eq.\ (\ref{ReducEq2}) to summation over
half of the unit cell $1\leq m\leq N_{l}$,
\begin{widetext}
\[
\sum_{m=-\infty}^{\infty}\!\mathcal{F}_{Mr}(v_{n}-v_{m},n-m)\!=\!\sum
_{m=1}^{N_{l}}\sum_{l=-\infty}^{\infty}\!\left[  \mathcal{F}_{Mr}(v_{n}%
-v_{m},n-m+lN)\!+\!\mathcal{F}_{Mr}(v_{n}+v_{m},n+m-1+lN)\right]
\]
\end{widetext}
Similar decomposition was made for the magnetic interaction energy
in Eq.\ (\ref{ReducedEn}).

We explored chain phase structures by numerically solving Eqs.\
(\ref{ReducEq}) with respect to the pancake row displacement
$v_{n}$ and regular phase distribution $\phi_{r,n}$ for different
values of the parameters $a$, $N=2N_{l}$, and $\alpha$. Numerical
computations were performed on two Linux workstations with 2GHz
AMD Athlon  processors and on the nodes of the Argonne computing
cluster ``Jazz'' (350 nodes, each with 2.4 GHz Pentium Xeon
processor). In the following sections we review the results of
these calculations.

\subsection{Stability of crossing configuration}%

\begin{figure}
[ptb]
\begin{center}\includegraphics[width=3.2in]%
{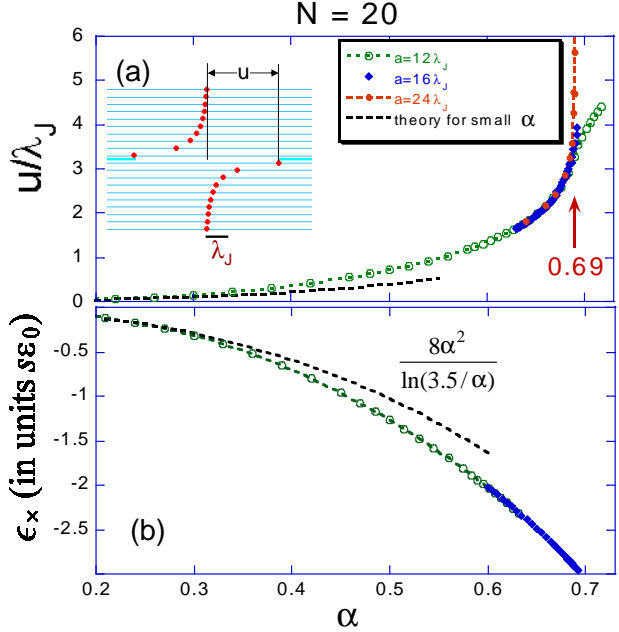}%
\caption{(a) The dependence of the maximum pancake displacement in the
crossing (defined in the inset) on the ratio $\alpha=\lambda/\lambda_{J}$ for
different periods $a$ in $x$ direction. Plot shows that crossing becomes
unstable near $\alpha=0.69$. Inset shows crossing configuration near
instability point. (b) The dependence of crossing energy on the ratio $\alpha
$. For comparison small $\alpha$ calculation is shown in both plots.}%
\label{Fig-CrossingResults}%
\end{center}
\end{figure}
The fundamental property of the crossing-chain state is the
structure of the crossing configuration of Josephson vortex and
pancake stack. In-plane currents of the Josephson vortex displace
the pancakes in the opposite directions above and below central
layers. Equilibrium displacements are the result of a balance
between the pulling forces of the Josephson vortex, which try to
tear the pancake stack apart, and magnetic coupling, which tries
to keep the stack aligned. The structure of the crossing
configuration can be calculated analytically in the regime of high
anisotropy $\lambda\ll\lambda_{J}$
\cite{CrossLatPRL99,JVpancPRB03} leading to the result
(\ref{eps_x}) for the crossing energy and to the maximum
displacement $u_{1}\approx2.2\lambda
^{2}/\left[\lambda_{J}\ln\left( 2\lambda_{J}/\lambda\right)
\right]$. This calculation is based on (i) quadratic approximation
for the magnetic tilt energy and (ii) assumption that the JV\
in-plane currents are not influenced much by pancake
displacements. The first assumption breaks down when $u_{1}$
approaches $\lambda$ and the second one breaks down when $u_{1}$
become comparable with $\lambda_{J}$. This means that the both
approximations break down as $\lambda$ approaches $\lambda_{J}$.

The stability of the crossing configuration has been addressed
recently by Dodgson \cite{DodgsonPhysC02} using the full magnetic
coupling energy but without taking into account modification of
the Josephson vortex by pancake displacements. The latter can not
be easily computed analytically. This calculation suggested that
the crossing configuration becomes unstable at
$\alpha\equiv\lambda/\lambda_{J}\approx1/2.86\approx0.35$. This
estimate seems to be in contradiction with the recent decoration
experiments \cite{TokunagaPRB03} where the isolated pancake stacks
sitting on the Josephson vortices have been observed in the
strongly overdoped BSCCO with the ratio $\alpha$ significantly
larger than this value.

To resolve this contradiction and find an accurate stability
criterion we studied numerically evolution of the isolated
crossing configuration with increasing ratio $\alpha$. For this
purpose we used the code, which calculates the chain structure in
Fig.\ \ref{Fig-CrossTiltChain}, for large values of the periods
$a$ and $N=2N_{l}$. Figure \ref{Fig-CrossingResults}a shows the
dependence of the maximum pancake displacement on the ratio
$\alpha$ for $N=20$ and different values of $a$. One can see that
the crossing configuration becomes unstable near $\alpha\approx
0.69$, which significantly exceeds a simple estimate in Ref.\
\onlinecite{DodgsonPhysC02}. The main reason for the extended
stability range is that the pancake displacements significantly
modify the structure of the Josephson vortex. This reduces forces
which pull the pancake stacks away and compensates for reduced
magnetic coupling restoring forces at large $u$. Finite-size
effect is only noticeable in vicinity of the instability. The
small-$\alpha$ calculation correctly predicts the maximum
displacement up to $\alpha\approx 0.35$ and underestimates it at
larger $\alpha$. Figure \ref{Fig-CrossingResults}b shows the
$\alpha$ dependence of crossing energy $\epsilon_{\times}$. It
demonstrates a rather regular behavior almost up to the
instability point. Similar to the maximum displacement, the
small-$\alpha$ calculation gives an accurate estimate for
$\epsilon_{\times}$ up to $\alpha\approx 0.35$. For higher
$\alpha$ an absolute value of $\epsilon_{\times}$ exceeds the
analytical estimate.

\subsection{Typical phase diagram within the range $0.4\lesssim\alpha
\lesssim0.5$: Phase transition from crossing to tilted chains with
decreasing pancake separation \label{Sec-PhaseDiag_sm}}

\begin{figure*}[ptb]
\begin{center}
\includegraphics[width=4.57in]{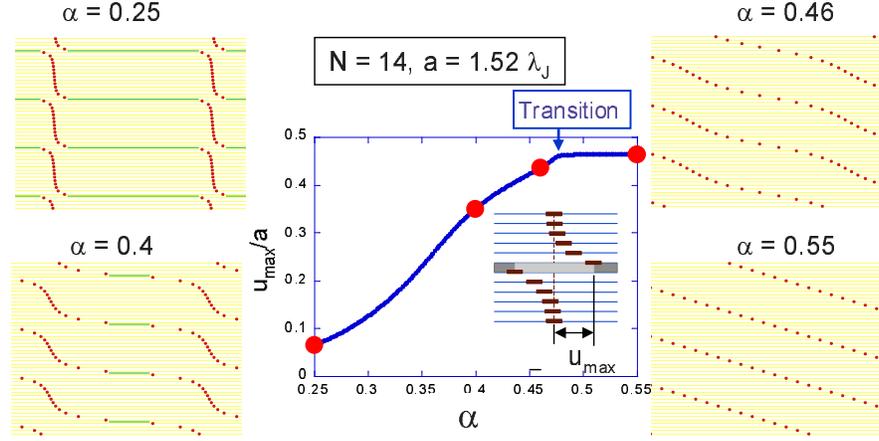}
\end{center}
\caption{\emph{Main plot} shows the dependence of the maximum
displacement(defined in the inset) divided by pancake separation
on the parameters $\alpha$ for $N=14$ and $a=1.52\lambda_{J}$. The
chain structures are illustrated at marked points. In the
configuration pictures circles show positions of the pancake
vortices and horizontal lines mark locations of the Josephson
vortices. One can see that the system evolves from weakly deformed
chain ($\alpha=0.25$) via strongly deformed chain ($\alpha=0.4$)
to modulated tilted chain ($\alpha=0.46$). The last structure
transforms via a second-order phase transition at $\alpha=0.48$
into tilted straight vortices.}
\label{Fig-ChainStructNl7ag1-52}%
\end{figure*}

At the first stage we studied the evolution of chain structures
with increasing $\alpha\equiv\lambda/\lambda_{J}$ for fixed
periods $a$ and $N$. As $\lambda$ increases with the temperature
and $\lambda_{J}$ is approximately temperature-independent,
increase of $\alpha$ corresponds to increase of the temperature in
real systems except that our calculations do not take into account
thermal fluctuations. For small values of $a$ and $N$ we found
that the chain structure evolves smoothly. An example of such
evolution is presented in Fig.\ \ref{Fig-ChainStructNl7ag1-52} for
$N=14$ and $a=1.52\lambda_{J}$. The main plot shows the dependence
of the maximum pancake displacement from the straight-stack
position $u_{max}/a$ (defined in the inset) on the parameter
$\alpha$. We will use this ratio to characterize the chain
structure throughout the paper. It changes from zero for straight
stacks to $(1-1/N)/2$ for tilted chains.  At small $\alpha$ a
weakly deformed crossing configuration is always realized (see the
structure at $\alpha=0.25$).  The pancake displacements grow with
increasing $\alpha$ and the chain evolves into strongly corrugated
configurations such as configuration for $\alpha=0.4$ in Fig.\
\ref{Fig-ChainStructNl7ag1-52}. With further increase of $\alpha$,
this structure smoothly transforms into modulated tilted lines
(see the structure for $\alpha=0.46$). Finally, the last structure
transforms via a second-order phase transition into the straight
tilted lines. For parameters used in Fig.\
\ref{Fig-ChainStructNl7ag1-52} this occurs at $\alpha=0.48$. The
plateau in the dependence $u_{max}(\alpha)$ above this value of
$\alpha$ corresponds to the maximum displacement $(1-1/N)/2$ in
the tilted chain.

\begin{figure}[ptb]
\begin{center}
\includegraphics[width=3.4in]{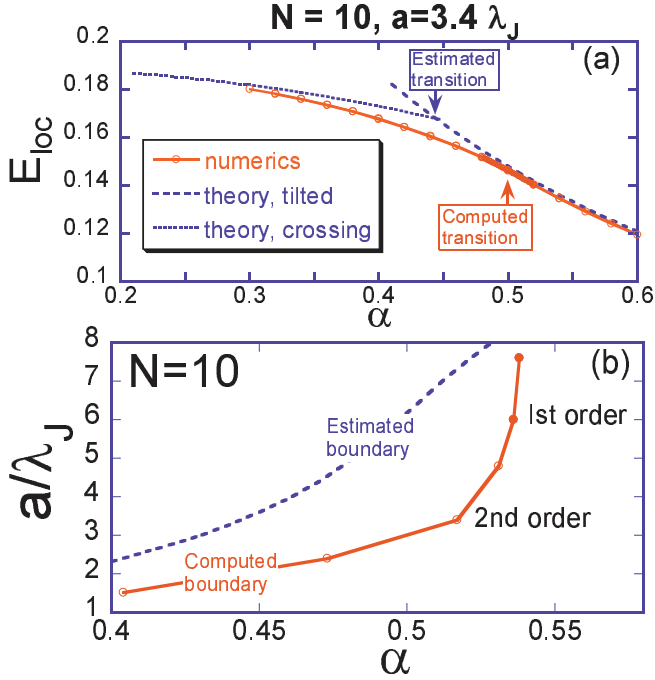}
\end{center}
\caption{(a) The dependence of the local part of energy of the
chain configuration on the ratio $\alpha$ for $N=10$ and
$a=3.4\lambda_{J}$. For comparison, we also show analytical
estimates for the crossing and tilted chain configuration. (b)
Phase diagram in the plane [pancake separation $a$]-[ratio
$\alpha$] for the fixed separation between JVs, $N=10$. The solid
line shows computed transition into the tilted chain state. The
dashed line shows location of the transitional region where the
energy of the crossing configuration is equal to
the energy of the tilted configuration. }%
\label{Fig-CompTheoryNl10}%
\end{figure}

To compare numerical and analytical calculations we plot in Fig.\
\ref{Fig-CompTheoryNl10}a the numerically computed
$\alpha$-dependence of the local energy (\ref{ReducedLocEn})
together with analytical estimates (\ref{Reduced_Eloc_est}) for
$N=10$ and $a=3.4\lambda_{J}$. One can see that the analytical
estimates accurately reproduce numerical results for the weakly
deformed crossing chain at $\alpha<0.35$ and for the tilted chain
at $\alpha>0.5$. However, in between the numerical study predicts
intermediate configurations with energies smaller than the
energies of the both limiting configurations. As a result, a
naively expected first-order phase transition is replaced by a
continuous transition occurring at significantly larger $\alpha$.
This behavior is quite general. We observed it within the broad
range of the periods, $6\lesssim N\lesssim20$, $2\lesssim
a/\lambda_{J}\lesssim5$ and the ratios
$0.4\lesssim\alpha\lesssim0.6$. In Fig.\ \ref{Fig-CompTheoryNl10}b
we compare location of the phase transition into the tilted-chain
state in the $a$-$\alpha$ plane for $N=10$ with location of the
transitional region defined by Eq.\ (\ref{Trans_Crit_sm}). The
computed transition line is always displaced from the estimated
boundary in the direction of larger $\alpha$. The transitional
region just marks the location of the intermediate strongly
deformed chain configurations. The observed continuous phase
transition indicates that tilted lines become unstable with
decrease of $\alpha$. It is known that an isolated vortex line in
anisotropic superconductors is unstable within some range of tilt
angles.\cite{ThompsMoorePRB97,BenkraoudaPRB96} We have to note
that the stability criterion of a chain is not identical to the
stability criterion of an isolated vortex line and requires
separate study. At large values of $a$ a continuous transition is
replaced by a first-order phase transition. However, as it was
discussed in Sec. \ref{Sec-Attrac}, due to the attractive
interaction between the pancake stacks, large separations may not
realize in equilibrium because $a$ is expected to jump from
infinity to the maximum equilibrium separation
$a_{m}$.

\begin{figure}[ptb]
\begin{center}
\includegraphics[width=3.4in]{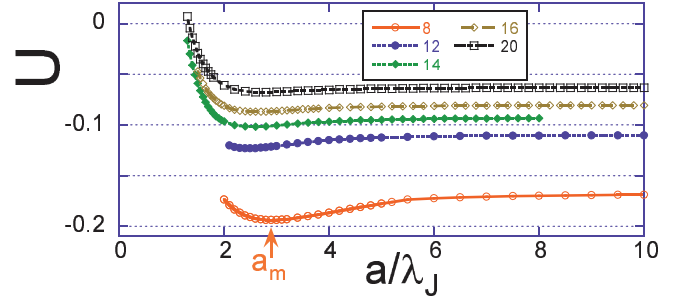}
\end{center}
\caption{The dependence of the excess pancake energy per stack in
units of $\varepsilon_0$, $U$, on the separation between pancakes
$a$ for $\alpha=0.5$ and different $N$. The position of minimum of
this energy corresponds to the maximum equilibrium separation
$a_{m}$
(marked for $N=8$ plot).}%
\label{Fig-am_al05}%
\end{figure}

To find location of $a_{m}$ in the phase space, we calculated the
evolution of chain structures with changing $a$ for fixed $\alpha$
and $N$. These calculations, of course, reproduce the chain
structures and location of the transition line of the previous
calculation. Following recipe of Sec. \ref{Sec-Attrac}, we
calculated the $a$-dependence of the excess pancake energy per
unit stack, $U (\tilde{a})\equiv\tilde {a}(E-E_{JV})$, and find
$a_{m}$ from the minimum location of this energy. Figure
\ref{Fig-am_al05} shows an example of these dependencies for
$\alpha=0.5$ and different $N$. Increasing $U(a)$ at $a>a_{m}$
implies attractive interaction between stacks at large
separations. We can see that for $\alpha=0.5$ the separation
$a_{m}$ weakly depends on $N$ and lies in between $2\lambda_{J}$
and $3\lambda_{J}$. Note that the $N$-dependence of the limiting
value of $U(a)$ at $a\rightarrow \infty$ reflects contribution
from the crossing energy of an isolated pancake stack with the JV
array.

\begin{figure}[ptb]
\begin{center}
\includegraphics[width=3.4in]{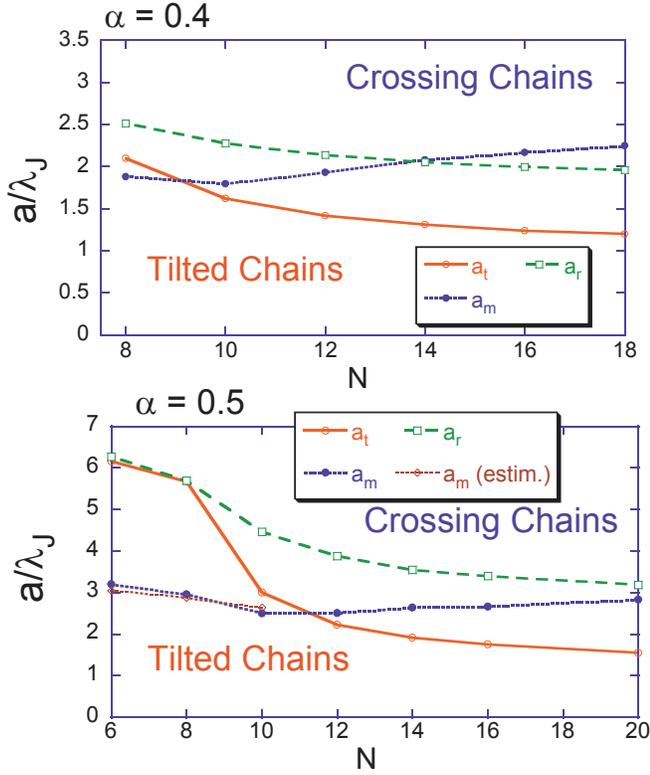}
\end{center}
\caption{Chain phase diagrams in the plane $a/\lambda_{J}-N$ for
two values of the ratio $\alpha$, 0.4 and 0.5. The solid line
indicate phase transition into the tilted-chain phase. The dotted
line shows the maximum equilibrium separation $a_{m}$. The dashed
line shows crossover ``reconnection'' line at which the maximum
displacement exceeds $a/4$ and weakly deformed crossing
configuration crosses over into the strongly corrugated
configuration. For $\alpha=0.5$ we also show $a_{m}$ for the
tilted-chain phase analytically calculated from
Eq.\ (\ref{a_m_tilted}).}%
\label{Fig-PhasDiagAl0405}%
\end{figure}

In Fig.\ \ref{Fig-PhasDiagAl0405} we present the chain phase
diagrams in the $a$-$N$-plane for two values of $\alpha$, $0.4$
and $0.5$. Solid lines show the phase transition into the
tilted-chain state (the dependence $a_{t}(N)$). One can see that
at larger $N$ the transition takes place at smaller $a$. With
increasing $\alpha$ this line moves higher meaning that the
tilted-chain state occupies larger area in the phase space. At
large $a$ weakly deformed chain configurations are realized,
similar to ones shown in Fig.\ \ref{Fig-ChainStructNl7ag1-52} for
$\alpha=0.25$. With decreasing $a$ the chain configuration crosses
over into a strongly corrugated state. To mark location of this
crossover we show in the phase diagrams by dashed lines the
pancake separation at which the maximum pancake displacement
$u_{max}$ reaches $a/4$ (plot $a_{r}(N)$). This crossover can be
viewed as \textquotedblleft reconnection\textquotedblright\ of
pancake stack segments. Dotted lines show locations of $a_{m}$. We
see that $a_{m}(N)$ line crosses the transition line meaning that
at small $N$ $a_{m}$ falls into the tilted-chain region and at
large $N$ it falls into the crossing-chains region. For
$\alpha=0.5$ we also show the analytical estimate for $a_{m}$ for
the tilted chain calculated from Eq.\ (\ref{a_m_tilted}). One can
see that it agrees very well with numerical calculations. For
$\alpha=0.5$ and $N\leq8$ the transition is of the first order.
However, the pancake separation at the transition lies above
$a_{m}$ meaning that it does not correspond to the ground state.

The obtained phase diagrams imply that at small tilting angle of
the field with respect to the $c$ axis (corresponding to small
$a$) the tilted chains have lower energy than the crossing chains.
This is similar to the situation at higher fields, in the dense
lattice, where the crossing-lattices state also is expected to
transform into the simple tilted lattice at small tilting angle of
the field.\cite{CrossLatPRL99}

\subsection{Typical phase diagram in the range $0.5\lesssim\alpha\lesssim
0.65$: Reentrant transition to kinked/tilted lines at small
concentration of pancake vortices}

At higher values of the ratio $\alpha$ a new qualitative feature
emerges in the phase diagram. When $\alpha$ exceeds the
characteristic value, a small $c$-axis field penetrates
superconductor in the form of kinks forming kinked vortex lines
(lock-in transition, see e.g., Refs.\
\onlinecite{IvlevMPL1991,Feinberg93,BLK}). The critical value of
$\alpha$ is determined by combination of numerical constants in
the pancake-stack and kink energies and it is given by $\alpha
_{c}=\exp C_{kv}$, where the constant $C_{kv}$ is defined after
Eq.\ (\ref{E_TVsingle_large}). To find the value of $\alpha_c$, we
calculated in Appendix \ref{App:KinkInter} the energy of tilted
chains at very large pancake separations $a$, which allowed us to
extract the energy of an isolated kink. This calculation gives
$C_{kv}\approx -0.81$ corresponding to $\alpha_c\approx 0.44$. It
is important to note that the critical value of $\alpha$ increases
with decrease of $N$, due to the increasing contribution of the
crossing energies to the total energy of crossing chain. Very
interesting behavior is expected when $\alpha$ is only slightly
larger than $\alpha_{c}$. The competing chain states have very
different interactions: deformed stacks attract and kinks repel
each other. Moreover, at the same value of the c-axis magnetic
induction, $B_{z}$, the kink separation is much smaller than the
stack separation and the absolute value of the kink interaction
energy is much larger than the interaction energy between deformed
stacks. As a consequence, with increasing $B_{z}$ the total energy
of the kinked lines rapidly exceeds the total energy of the
crossing chain and the system experiences a first-order phase
transition into the crossing-chain state. Due to the attractive
interaction between the pancake stacks, the pancake/kink
separation at which the energy curves cross does not give the
equilibrium separation for the crossing chain and the stack
separation jumps at the transition to a value slightly smaller
than the maximum equilibrium separation $a_{m}$. This means that
the phase transition is accompanied by jump of pancake density and
magnetic induction, $B_{z}$.

\begin{figure}[ptb]
\begin{center}
\includegraphics[clip,width=3.4in]{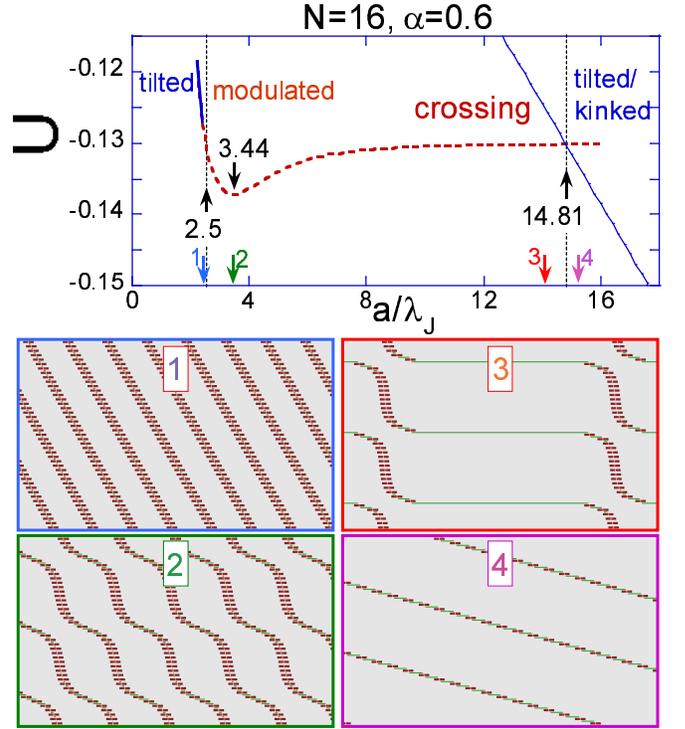}
\end{center}
\caption{The dependence of the excess pancake energy per stack on
pancake separation $a$ for $N=16$ and $\alpha=0.6$. The two
branches correspond to the two different starting states at large
$a$, crossing chain and kinked lines. One can see that the kinked
lines have lower energy at very large $a$, $a>14.81$. Crossing
chain smoothly transforms back into tilted chain with decrease of
$a$. The transformation is completed at a second-order phase
transition point at $a_{t}=2.5\lambda_{J}$. Chain configurations
at four points marked by arrows are shown below. The evolution of
chain configuration along this energy curve is also illustrated by
an animation.\cite{AnimationN16al06}}
\label{Fig-adEN16Al0_6}%
\end{figure}%
This behavior was confirmed by numerical calculations. Figure
\ref{Fig-adEN16Al0_6} shows a plot of the dependence of the
pancake energy $U$ on the pancake separation for $N=16$ and
$\alpha=0.6$. This dependence has two branches, corresponding to
the two different starting states at large $a$, crossing chain and
kinked lines. This branches cross at $a=14.8\lambda_{J}$ and the
the kinked vortex lines have smaller energy at larger $a$. The
variations of $U$ at large $a$ occur due to the interaction energy
and one can see that the kink interaction energy is much larger
than the interaction energy of the deformed stack in the crossing
chain. With further decrease of $a$, the crossing chain smoothly
transforms into the tilted chain, as it was described in the
previous Section. The second-order phase transition for these
parameters takes place at $a\approx 2.5\lambda_{J}$ somewhat
smaller than the maximum equilibrium separation
$a_{m}\approx3.44\lambda_{J}$.%
\begin{figure}[ptbptb]
\begin{center}
\includegraphics[width=3.2in]{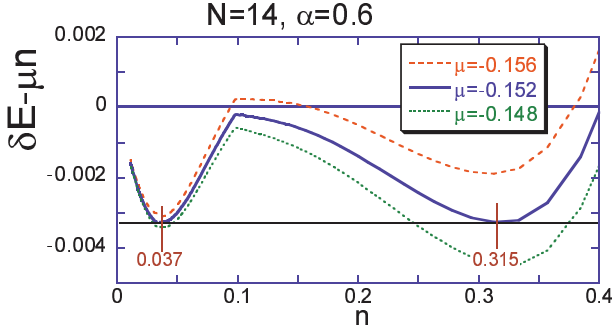}
\end{center}
\caption{The density dependence of the pancake part of the
thermodynamic potential per unit area, $\delta E-\mu n$ (in units
of $\epsilon _{0}/\lambda_{J} $) at different chemical potentials
$\mu$ corresponding to different values of the magnetic field
strength for $N=14$ and $\alpha=0.6$. Kinks in the curves separate
regions of tilted/kinked lines (low $n$) and crossing chains (high
$n$). The equilibrium density in units of $1/\lambda_{J}$ is given
by the global minimum of this energy. One can see that at $\mu
\approx-0.152$ the system experiences a
first-order phase transition with very large density jump.}%
\label{Fig-dE-munN16Al0_6}%
\end{figure}%

The pancake separation (or pancake density) in the chain can not
be directly fixed in experiment. Instead, the magnetic field
strength, $H_z$, fixes the chemical potential $\mu_{H}$,
$\mu_{H}=\Phi_{0} H_z/(4\pi)$, and the equilibrium density is
determined by the global minimum of the thermodynamic potential
$G(n)=E(n)-\mu_{H} n$. To find evolution of density with
increasing chemical potential, we plot in Fig.\
\ref{Fig-dE-munN16Al0_6} the density dependencies of the reduced
thermodynamic potential, $\delta E -\mu n$ for different $\mu$ and
representative parameters $N=14$ and $\alpha=0.6$.  As the energy
of isolated stacks is subtracted in $\delta E$, the dimensionless
chemical potential is shifted with respect to its bare value and
it is related to the magnetic field strength
as%
\[\mu=\frac{\Phi_{0}(H_z-H_{c1})}{4\pi \epsilon_0}\]%
where $H_{c1}$ is the lower critical field for $\mathbf{H}\| c$.
We find that for selected parameters the transition takes place at
$\mu=\mu_{t}=-0.152$. At $\mu<\mu_{t}$ the global minimum falls
into the region of kinked lines and at $\mu>\mu_{t}$ it jumps into
the region of crossing chain. Note that the density value, at
which the energy curves cross (kinks in the lines near $n= 0.1$ in
Fig.\ \ref{Fig-dE-munN16Al0_6}), is always larger than the lower
density from which the jump to the high-density state takes place.
At the transition, the density jumps almost ten times, from
$0.037/\lambda _{J}$ to $0.315/\lambda_{J}$.

\begin{figure}[ptb]
\begin{center}
\includegraphics[width=3.2in]{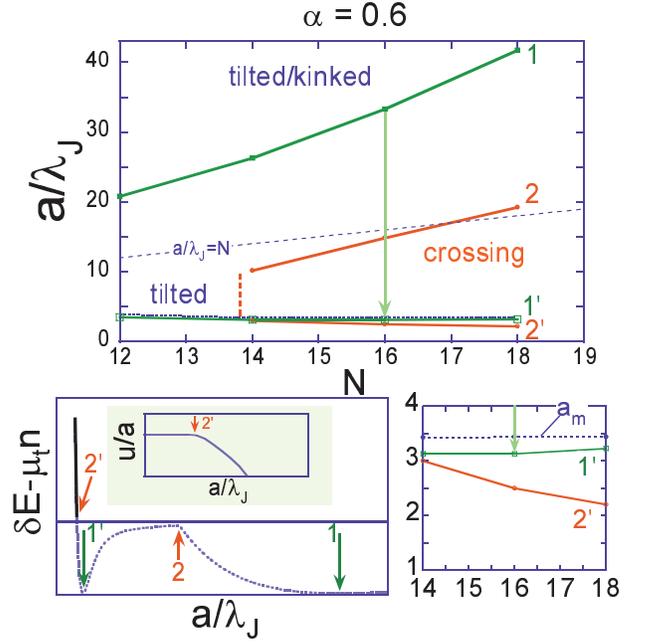}
\end{center}
\caption{\emph{The upper plot} shows the phase diagram in the
$N$-$a$ plane for $\alpha=0.6$. The left plot in the lower plat
illustrates meanings of the phase lines using the plot of the
thermodynamic potential $\delta E-\mu n$ vs $a/\lambda_{J}$ at the
transition point (main plot) and the maximum displacement $u$
divided by lattice spacing $a$ vs $a/\lambda_{J}$ (inset). The
lines $1$ and $1^{\prime}$ correspond to the two limiting pancake
separations at the transition point between which the jump occurs.
The line 2 indicates crossing of the energy curves for the kinked
and crossing chain. The line $2^{\prime}$ shows the position of a
continuous transition into the tilted chain (illustrated by the
inset in the lower panel). Dotted line slightly above
$1^{\prime}$-line shows the position of the maximum equilibrium
separation $a_{m}$. We also show the crossover line
$a/\lambda_{J}=N$ above which well-defined kinks appear. The right
plot in the lower part shows blowup of the phase diagram above the
$2^{\prime}$ line.}
\label{Fig-PhaseDiagAl0_6}%
\end{figure}%
The numerically obtained phase diagram in the $N$-$a$ plane for
$\alpha=0.6$ is shown in the upper panel of Fig.\
\ref{Fig-PhaseDiagAl0_6}. The plots in the lower left panel, the
thermodynamic potential at the transition point and the maximum
displacement versus $a/\lambda_{J}$, illustrate definitions of
different lines in the phase diagram. The lines $1$ and
$1^{\prime}$ show the limiting pancake separations at the
first-order transition between which the jump takes place. When
the chemical potential is fixed by external conditions, the area
between these lines is bypassed in equilibrium. At large $N$ the
jump takes place from the kinked-lines state into the strongly
corrugated configuration. This configuration transforms into the
tilted chain with further decrease of $a$ via continuous
transition shown by the line $2^{\prime}$. Below $N=14$ only
tilted chains realize, but the density jump still exists. The
upper separation grows approximately proportional to $N$ while the
lower separation slowly decreases with $N$ and lies slightly below
the maximum equilibrium separation $a_{m}$ shown by dotted line.
This means that the relative density jump increases with $N$. The
line $2$ shows the position of the crossing of the energy curves
for the two states.

\begin{figure}
[ptbptb]
\begin{center}
\includegraphics[width=3.4in]
{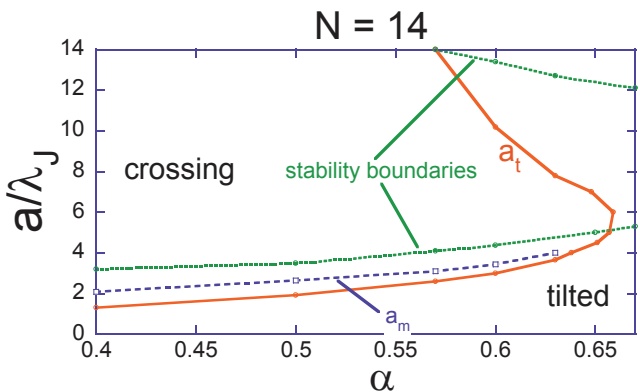}%
\caption{Phase diagram in $a$-$\alpha$ plane for $N=14$. Solid
nose-shaped line shows the phase transition into the tilted-chain
state. Dashed line shows location of the maximum equilibrium
separation, which terminates at some point. We also show by dotted
line location of the stability boundaries obtained from inflection
points of the dependencies $U(a)$ (see
criterion(\ref{StabCrit})).}
\label{Fig-PhDiagN14}%
\end{center}
\end{figure}

\begin{figure}
[ptbptbptb]
\begin{center}
\includegraphics[ width=3.4in]{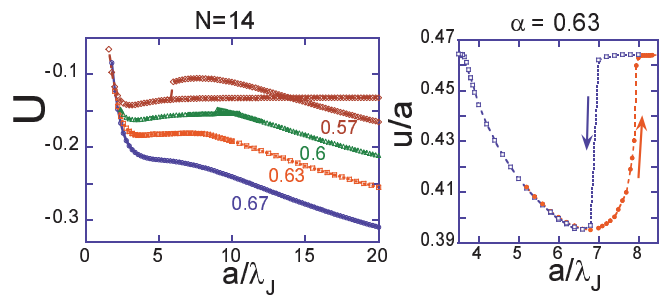}
\caption{\emph{Left panel:} The dependencies of the excess pancake
energy per stack $U$ on separation $a$ for $N=14$ and different
values of $\alpha$. With increasing $\alpha$ the first-order
transition from tilted to crossing chain vanishes and the crossing
chain do not realize at all. Also the dependence $U(a)$ becomes
monotonic which corresponds to termination of $a_m(\alpha)$ line
in Fig.\ \ref{Fig-PhDiagN14}. \emph{Right panel:} The dependence
of the relative maximum displacement $u$ on separation $a$ for
$N=14$ and $\alpha =0.63$. The maximum value of $u$, corresponding
to tilted chains, is given by $(1-1/14)/2\approx 0.464$. One can
see that there is a range of $a$ where the chains become only
slightly modulated. } \label{Fig-adEN14}
\end{center}
\end{figure}

It is also instructive to examine the phase diagram for fixed
$c$-axis period $N$, in the $a$-$\alpha$ plane. As $\alpha$
increases with temperature, this diagram to some extent describes
the temperature evolution of the chain structure. Figure
\ref{Fig-PhDiagN14} shows such diagram for $N=14$. One can see
that the phase transition line is reentrant, within some range of
$\alpha$ the tilted chains are realized both for small and large
$a$. Above some critical value of $\alpha$ ($0.66$ for $N=14$)
only straight tilted lines exist in whole range of $a$. This
maximum value increases with $N$. For somewhat smaller values of
$\alpha$, there is narrow range of $a$ where the chains become
slightly modulated. This is illustrated in the plot of the maximum
displacement $u$ for $\alpha =0.63$ in the right panel of Fig.\
\ref{Fig-adEN14}. The maximum equilibrium separation line
$a_{m}(\alpha)$ terminates at certain value of $\alpha $, above
which the dependence $U(a)$ becomes monotonic. We also show in
this diagram the stability boundaries extracted from the
inflection points in the $U(a)$ dependencies (see criterion
(\ref{StabCrit}) and related discussion). Origin of such phase
diagram can be better understood by studying the dependencies
$U(a)$ at different $\alpha$ shown in the left panel of Fig.\
\ref{Fig-adEN14}. One can see that the first-order transition
vanishes above certain value of $\alpha$ and the dependence $U(a)$
becomes monotonic close to this value of $\alpha$. On the other
hand, even after becoming monotonic, this dependence still has two
inflection points bounding unstable region in some range of $a$.
Existence of such unstable region means that the pancake-density
jump with increasing the c-axis external field persists in the
region where only tilted chains exist. This jump is closely
related (but not identical) with the jump of the tilt angle of the
vortex line at the lower critical field with increasing tilt angle
of the external field.\cite{KlemmClemPRB80,MorgadoPhysC01}

The vortex chains in BSCCO at small concentrations of pancakes
have been studied by the scanning Hall probe microscopy in Ref.\
\onlinecite{GrigNat01}. It was found that at very small
concentration of pancakes the chains are magnetically homogeneous
and separate pancake stacks are not resolved. When the external
field exceeds certain critical value, crystallites of the pancake
stacks are suddenly formed along the chains and the flux density
in crystallites approximately ten times higher then the flux
density in homogeneous chains. The [kinked lines]-[crossing
chains] first-order phase transition provides a very natural
explanation for this observations.

\section{Conclusions}

In conclusion, we investigated numerically and analytically the
phase diagram of an isolated vortex chain in layered
superconductors. In the region where Josephson and magnetic
coupling are approximately equal, we found a very rich behavior.
The crossing chains typically transform into tilted chains with
decreasing pancake separation via formation of intermediate
strongly deformed configurations and a continuous phase
transition. When the relative strength of the Josephson coupling
exceeds some typical value, the phase diagram becomes reentrant.
At a very small c-axis field, tilted chains are realized in which
the vortex lines have the kinked structure. With increasing c-axis
field these low-density tilted chains transform via a first-order
phase transition into strongly-deformed crossing chains. This
transition is accompanied by a large jump of the pancake-vortex
density. With further increase of the field these crossing chains
transforms back into the tilted chains via a second-order
transition.

An important feature of real BSCCO which is not taken into account
in this paper is the thermal vortex fluctuations. We expect that
fluctuations will not change qualitatively the described behavior,
especially the strong first-order phase transition, but may
significantly change locations of the phase transitions in the
phase space.

Finally, we briefly overview the relevant field scales. In the
range of studied JV separations, $N$, from $10$ to $20$, and for
$\gamma=500$, the in-plane field, $B_x=2\Phi_0/(\sqrt{3}\gamma s^2
N^2)$, varies in the range from 200 to 50 Gauss and the in-plane
separation between Josephson vortices, $c_y=(\sqrt{3}/2)\gamma s
N$, varies from 6.75 to 13.5 $\mu$m. The typical density jump for
$N=16$ and $\alpha=6$ in Fig.\ \ref{Fig-PhaseDiagAl0_6}
corresponds to jump if the average c-axis induction,
$B_z=\Phi_0/ac_y$, from 0.12 to 1.2 Gauss. Such jump is usually
difficult to notice in the global magnetization measurements.
Therefore, a rich spectrum of transformations discussed in this
paper takes place in the range of very small c-axis magnetic
induction, not exceeding few gauss. Taking a typical value
$\lambda$ for BSCCO at 80K, as 0.4 $\mu$m, we estimate for the
same parameters that the maximum c-axis field in the chain center,
$B_{z0}=\Phi_0/a\lambda$, jumps from 3.3 Gauss to 33 Gauss.

\section{Acknowledgements}

I would like to thank M.\ Dodgson, A.\ Grigorenko, S.\ Bending,
and V.\ Vlasko-Vlasov for useful discussions. I also gratefully
acknowledge use of ``Jazz'', a 350-node computing cluster operated
by the Mathematics and Computer Science Division at Argonne
National Laboratory as part of its Laboratory Computing Resource
Center. This work was supported by the U.\ S.\ DOE, Office of
Science, under contract \# W-31-109-ENG-38.

\appendix

\section{Kink interaction energy of a single line \label{App:KinkInter}}

We compute the kink interaction energy within London
approximation. As the main contribution to this energy comes from
the regions away from the JV and kink cores, one can expect that
the London approach gives a very good approximation
of the interaction energy. Shape of the kinked line is given by%
\begin{align*}
\mathbf{R}(X)  &  =\left(  X,0,u(X)\right)  ,\\
u(X)  &  =ns,\ (n-1/2)L<X<(n+1/2)L.
\end{align*}
From the general formula (\ref{LondonVortEner}) we obtain the
total energy of the
kinked line in London approximation%
\begin{widetext}
\begin{equation}
\varepsilon_{\mathrm{kl}}=\frac{\Phi_{0}^{2}}{8\pi L_{x}}\int\frac
{d^{3}\mathbf{k}}{(2\pi)^{3}}\int dX\int dX^{\prime}\frac{(1+\lambda^{2}%
k^{2})+(1+\lambda_{c}^{2}k^{2})\frac{du}{dX}\frac{du}{dX^{\prime}}}%
{(1+\lambda^{2}k^{2})\left(  1+\lambda^{2}k_{z}^{2}+\lambda_{c}^{2}%
k_{\parallel}^{2}\right)  }\exp(\imath k_{x}(X-X^{\prime})+\imath
k_{z}(u-u^{\prime})) \label{En_kl_Lon}%
\end{equation}
To separate the kink interaction energy one has subtract from this
expression the energies of Josephson vortices,
$\varepsilon_{JV}^{L}$, and kinks,
$\varepsilon_{\mathrm{kink}}^{L}$, in London approximation.
Integration over $X$, $X^{\prime}$, and $k_{x}$ leads to the
following expression for the total kink contribution to energy,
$\varepsilon_{\mathrm{k}}=\varepsilon
_{\mathrm{kl}}-\varepsilon_{JV}^{L}$%
\[
\varepsilon_{\mathrm{k}}    =\frac{\Phi_{0}^{2}}{8\pi}\int\frac{dk_{y}dk_{z}%
}{(2\pi)^{2}}\left[  \left(
\frac{1}{1+\lambda^{2}k_{z}^{2}}-\frac
{1}{1+\lambda^{2}k_{z}^{2}+\lambda_{c}^{2}k_{y}^{2}}\right)  \frac{\sinh g}%
{g}\frac{1-\cos\left(  k_{z}s\right)  }{\cosh g-\cos\left(
k_{z}s\right) }  +\frac{s^{2}/\lambda^{2}}{2\left(
1+\lambda^{2}k_{z}^{2}\right) }\frac{\sinh g_{1}/g_{1}}{\cosh
g_{1}-\cos k_{z}s}\right]
\]
with%
\begin{align*}
g  &  =L\sqrt{\lambda_{c}^{-2}+\gamma^{-2}k_{z}^{2}+k_{y}^{2}},\\
g_{1}  &  =L\sqrt{\lambda^{-2}+k_{y}^{2}+k_{z}^{2}}.
\end{align*}
To separate kink interaction we have to subtract the contribution
coming from isolated kinks, i.e., the term which behaves as $1/L$
at $L\rightarrow\infty$.
\[
\varepsilon_{\mathrm{kink}}^{L}=\frac{\Phi_{0}^{2}}{8\pi}\int\frac
{dk_{y}dk_{z}}{(2\pi)^{2}}\left[  -\frac{1-\cos\left(  k_{z}s\right)
}{g\left(  1+\lambda^{2}k_{z}^{2}+\lambda_{c}^{2}k_{y}^{2}\right)  }%
+\frac{s^{2}}{2\lambda^{2}\left(  1+\lambda^{2}k_{z}^{2}\right)  }\left(
\frac{1}{g_{1}}+\frac{\lambda^{2}k_{z}^{2}}{g}\right)  \right]  .
\]
This gives the following result for the kink interaction energy%
\begin{align}
\varepsilon_{\mathrm{ki}}  &  =\frac{\Phi_{0}^{2}}{8\pi}\int\frac{dk_{y}%
dk_{z}}{(2\pi)^{2}}\left[  \left(  \frac{1}{1+\lambda^{2}k_{z}^{2}}-\frac
{1}{1+\lambda^{2}k_{z}^{2}+\lambda_{c}^{2}k_{y}^{2}}\right)  \frac
{1-\cos\left(  k_{z}s\right)  }{g}\frac{\cos k_{z}s-\exp\left(  -g\right)
}{\cosh g-\cos k_{z}s}\right. \label{E_ki}\\
&  \left.  +\frac{s^{2}/\lambda^{2}}{2\left(  1+\lambda^{2}k_{z}^{2}\right)
}\frac{1}{g_{1}}\frac{\cos\left(  k_{z}s\right)  -\exp\left(  -g_{1}\right)
}{\cosh g_{1}-\cos\left(  k_{z}s\right)  }\right]  ,\nonumber
\end{align}
which has to be evaluated in the limit $L>\gamma s$. The main contribution is
coming from the first term in square brackets. If we keep only this term, than
the kink interaction energy can be reduced to the following form%
\[
\varepsilon_{\mathrm{ki}}\approx\frac{\gamma\Phi_{0}^{2}s^{2}}{8\pi\lambda
^{2}L^{2}}J(L/\lambda_{c}).
\]
with%
\begin{align*}
J(r)  &  =\int\frac{dp_{y}dp_{z}}{(2\pi)^{2}}\left(  \frac{1}{1+p_{z}^{2}%
}-\frac{1}{1+p^{2}}\right)  \frac{p_{z}^{2}}{\sqrt{1+p^{2}}}\frac{r}%
{\exp\left(  r\sqrt{1+p^{2}}\right)  -1},\\
p^{2}  &  \equiv p_{y}^{2}+p_{z}^{2}%
\end{align*}
\end{widetext}
In the practically interesting case $r=L/\lambda_{c}\ll1$ the
integral $J(r)$
can be evaluated as%
\[
J\approx\frac{1}{4\pi}\left(  \ln\left(  \frac{1}{r}\right)  -\frac{3}%
{2}\right)  ,
\]
giving the main result for the kink interaction energy (\ref{En_ki_result}).

The second term in square brackets of Eq.\ (\ref{E_ki}) represents magnetic
coupling contribution to the kink interactions. We calculated this
contribution in the two limiting cases%
\[
\varepsilon_{\mathrm{ki}}^{(2)}\approx%
\genfrac{\{}{.}{0pt}{}{\frac{s^{2}\varepsilon_{0}}{L^{2}}\text{, for }%
L\ll\lambda}{\frac{s^{2}\varepsilon_{0}}{\lambda L}\exp(-L/\lambda)\text{, for
}L\ll\lambda}%
\]
As we can see, it does give a very small contribution to the total kink
interaction energy.

\section{Derivation of the interaction energy of tilted vortices.
\label{App:InterTilted}}

The interaction potential between two straight tilted vortices per
unit length along c-axis
separated by distance $R$ in the tilt direction ($x$ axis) is given by%
\begin{widetext}
\[
U_{i}(R)=\frac{\Phi_{0}^{2}}{4\pi}\int\frac{dk_{x}dk_{y}}{(2\pi)^{2}}\frac
{\nu^{2}+(1+\lambda_{c}^{2}(k_{\perp}^{2}+\nu^{2}k_{x}^{2}))/(1+\lambda
^{2}(k_{\perp}^{2}+\nu^{2}k_{x}^{2}))}{1+\lambda^{2}\nu^{2}k_{x}+\lambda
_{c}^{2}k_{\perp}^{2}}\exp(ik_{x}R)
\]
\end{widetext}%
This formula works also in the regime of kinked vortex lines
$\nu>\gamma$. The kink structure of the lines starts to influence
interaction between them when kink separation $L$ exceeds
$c_{z}\gamma/2\pi$ corresponding to the condition
$\nu>N\gamma/2\pi$. Integrating over $k_{y}$ we obtain
$U_{i}(R)=\int
(dk/2\pi)\cos\left(  kR\right)  U_{i}(k)$ with%
\begin{equation}
U_{i}(k)=2\pi\varepsilon_{0}\lambda\!\sum_{j=1}^{3}g_{j}(k) \label{IntEn}%
\end{equation}
with
\begin{align*}
g_{1}(k)  &  =\frac{\nu^{2}\lambda^{2}+\lambda_{c}^{2}}{\lambda_{c}\left(
1+(\nu^{2}\lambda^{2}+\lambda_{c}^{2})k^{2}\right)  ^{1/2}},\\
g_{2}(k)  &  =-\frac{\lambda_{c}}{\left(  1+(\nu^{2}\lambda^{2}+\lambda
_{c}^{2})k^{2}\right)  ^{1/2}\left(  1+\nu^{2}\lambda^{2}k^{2}\right)  },\\
g_{3}(k)  &  =\frac{\lambda}{\left(  1+\lambda^{2}(1+\nu^{2})k^{2}\right)
^{1/2}\left(  1+\nu^{2}\lambda^{2}k^{2}\right)  }.
\end{align*}
Interaction energy of the chain per unit area $E_{TV}^{i}$ can be represented
as%
\begin{align*}
E_{TV}^{i}  &  =\frac{1}{2a}\sum_{n\neq0}\int\frac{dk}{2\pi}U_{i}%
(k)\exp(ikna)\\
&  =\frac{1}{2a^{2}}\sum_{m=-\infty}^{\infty}U_{i}(k_{m})-\frac{1}{2a}%
\int\frac{dk}{2\pi}U_{i}(k)
\end{align*}
with $k_{m}=2\pi m/a$, or%
\[
E_{TV}^{i}=\sum_{j=1}^{3}E_{TV,j}^{i}%
\]
with
\[
E_{TV,j}^{i}=\frac{\pi\varepsilon_{0}}{a^{2}}\sum_{m=-\infty}^{\infty}\left(
g_{j}(k_{m})-\int_{-\pi/a}^{\pi/a}\frac{adk}{2\pi}g_{j}(k_{m}+k)\right)
\]
In the limit $a\ll2\pi\sqrt{\nu^{2}\lambda^{2}+\lambda_{c}^{2}}$
the first term, $j=1$, can be evaluated as
\begin{align*}
E_{TV,1}^{i}&\approx\frac{\varepsilon_{0}}{a}\left[
\frac{\pi\left( \nu ^{2}\lambda^{2}+\lambda_{c}^{2}\right)
}{\lambda_{c}a}\right.\\
&\left.-\frac{\sqrt{\nu ^{2}+\gamma^{2}}}{\gamma}\left( \ln\left(
\frac{4\pi\sqrt{\nu^{2}\lambda ^{2}+\lambda_{c}^{2}}}{a}\right)
-\gamma_{E}\right)  \right]
\end{align*}%
Using integral
\[
\int_{0}^{\infty}\frac{dk}{\sqrt{1+a^{2}k^{2}}(1+b^{2}k^{2})}\underset{a>b}%
{=}\frac{1}{\sqrt{a^{2}-b^{2}}}\ln\frac{\sqrt{a^{2}-b^{2}}+a}{b},
\]
other two terms are calculated in the limit $c\ll2\pi\lambda$ as
\begin{align*}
E_{TV,2}^{i}&+E_{TV,3}^{i}\approx\frac{\varepsilon_{0}}{a}\left[
-\frac
{\pi\lambda_{c}}{a}+\frac{\pi\lambda}{a}+\ln\frac{\gamma+\sqrt{\nu^{2}%
+\gamma^{2}}}{1+\sqrt{1+\nu^{2}}}\right.\\
&\left.-\zeta(3)\left(  \frac{c}{2\pi\lambda
}\right)  ^{2}\left(  \frac{\gamma}{\sqrt{\nu^{2}+\gamma^{2}}}-\frac{1}%
{\sqrt{1+\nu^{2}}}\right)  \right]
\end{align*}%
The last term is small and will be dropped in further
calculations. Collecting terms, we finally obtain for the total
interaction energy (\ref{EnTiltInt}).

\section{Energy of isolated kink \label{Sec-Kink}}

\begin{figure}[ptb]
\begin{center}
\includegraphics[width=3.4in]{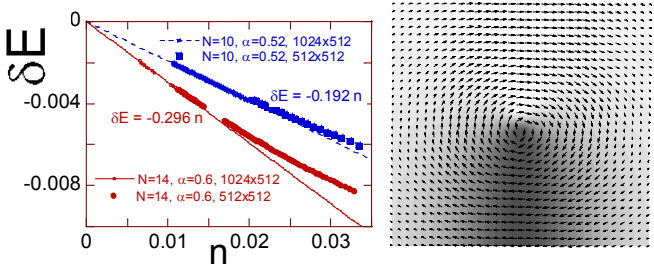}%
\caption{\emph{Left panel} shows plot of the pancake part of the
chain energy per unit area in units $\varepsilon_{0}/\lambda_{J}$,
$\delta E$, vs pancake density $n$ at very small $n$. Slope of
this energy at $n\rightarrow0$ determines the energy of an
isolated kink and the numerical constant $C_{kv}$. Calculations
were made for two sets of parameters, $(N=10,\alpha=0.52)$ and
$(N=14,\alpha =0.6)$, and for two system sizes for each set.
\emph{Right picture} illustrates the numerically calculated kink
structure. Arrows show in-plane currents and gray level plot codes
distribution of the cosine of phase difference between the
neighboring layers (dark area below corresponds to the Josephson
vortex.)} \label{Fig-kinkEn}
\end{center}
\end{figure}
To find the energy of an isolated kink, we calculated energy of
tilted lines in the regime when kink separation $L=a/N$
significantly exceeds the Josephson length. Numerically, this is a
challenging task because the kink interaction energy decays slowly
with increasing $L$ meaning that one has to go to very large
values of $a$. To maintain sufficient accuracy, one has to use
large number of grid points in $x$ direction. As follows from Eq.\
(\ref{E_TVresult_lrg}), the pancake part of energy vanishes
linearly at small kink concentrations $n$. In reduced units, we
define this energy as $\delta\tilde{E}=(\lambda
_{J}/\varepsilon_{0})(E_{TV}-E_{PV}^{s}-E_{JV})$ and from
Eq.\ (\ref{E_TVresult_lrg}) we have%
\[
\delta\tilde{E}\approx n\left(  \ln\frac{1}{\alpha}+C_{kv}\right)
\]
Plots of this energy are shown in Fig.\ \ref{Fig-kinkEn} for two
sets of parameters, $(N=10,\alpha=0.52)$ and $(N=14,\alpha=0.6)$.
From linear fits at small $n$ we obtain estimates
$C_{kv}\approx-0.192-\ln(1/0.52)\approx0.846$ for the first set
and $C_{kv}\approx-0.296-\ln(1/0.6)\approx0.807$ for the second
set. If we use the last constant, corresponding to the larger
system size, we obtain the estimate $C_{k}\approx-0.31$ for the
constant in the kink energy within Ginzburg-Landau theory
(\ref{kink_Ener}). This is somewhat smaller than the value $-0.17$
reported in Ref.\ \onlinecite{Kinkwalls}. The difference is most
probably due to finite-size effects.


\begin{thebibliography}{99}
\bibitem{note:KinkWalls}The condition $\nu> N\gamma/2\pi$ means that the
separation between kinks $L$ exceeds $c\gamma/2\pi$. In this
region the kink interaction energy is strongest for the kinks in
the same $x$-$z$-plane. This planar array of kinks has been called
``kink wall'' in Ref.\ \onlinecite{Kinkwalls}. Interaction between
the neighboring kink walls decays exponentially at large $L$.
\end{thebibliography}
\end{document}